\documentclass[twocolumn,twoside]{article} 

\usepackage{graphicx} %
\usepackage{epstopdf}

\usepackage{latexsym}
\usepackage{amsfonts}
\usepackage{amsmath}
\usepackage{makeidx}
\usepackage{amssymb}
\usepackage{enumerate}
\usepackage{fancybox}
\usepackage{marginfix}

\usepackage{caption}
\usepackage{comment}

\usepackage{geometry}
\usepackage{fancyhdr}
\let\OLDthebibliography\thebibliography
\renewcommand\thebibliography[1]{
  \OLDthebibliography{#1}
  \setlength{\parskip}{0pt}
  \setlength{\itemsep}{0pt plus 0.3ex}
}

\newif\ifen
\newif\ifjp

\newcommand{\en}[1]{\ifen#1\fi}

\entrue

\newcommand{\dbar}{{d\mkern-7mu\mathchar'26\mkern-2mu}}
\renewcommand\thefootnote{\arabic{footnote}}

\begin{document}

\en{ \title{The principles of adaptation in organisms and machines II: Thermodynamics of the Bayesian brain} }

\en{ \author{Hideaki Shimazaki} }

\en{ \date{Graduate School of Informatics, Kyoto University} }

\twocolumn[
  \begin{@twocolumnfalse}
    \maketitle
    \begin{abstract}

\en{This article reviews how organisms learn and recognize the world through the dynamics of neural networks from the perspective of Bayesian inference, and introduces a view on how such dynamics is described by the laws for the entropy of neural activity, a paradigm that we call thermodynamics of the Bayesian brain. The Bayesian brain hypothesis sees the stimulus-evoked activity of neurons as an act of constructing the Bayesian posterior distribution based on the generative model of the external world that an organism possesses. A closer look at the stimulus-evoked activity at early sensory cortices reveals that feedforward connections initially mediate the stimulus-response, which is later modulated by input from recurrent connections. Importantly, not the initial response, but the delayed modulation expresses animals' cognitive states such as awareness and attention regarding the stimulus. Using a simple generative model made of a spiking neural population, we reproduce the stimulus-evoked dynamics with the delayed feedback modulation as the process of the Bayesian inference that integrates the stimulus evidence and a prior knowledge with time-delay. We then introduce a thermodynamic view on this process based on the laws for the entropy of neural activity. This view elucidates that the process of the Bayesian inference works as the recently-proposed information-theoretic engine (neural engine, an analogue of a heat engine in thermodynamics), which allows us to quantify the perceptual capacity expressed in the delayed modulation in terms of entropy.}
\vspace{1em}
   \end{abstract}
\end{@twocolumnfalse}
]

{
  \renewcommand{\thefootnote}%
    {\fnsymbol{footnote}}
  \footnotetext[1]{Address: Yoshida-honmachi, Sakyo-ku, Kyoto 606-8501, Japan; Email: h.shimazaki@i.kyoto-u.ac.jp; URL: https://www.neuralengine.org}
}

\maketitle

\en{ \section{Introduction}\label{sec:prologue} }

\en{Living organisms learn the world, recognize the current situation, and then choose their actions for their survival. These processes are underpinned by the nonlinear dynamics of the network activity in nervous systems. Neuronal mechanisms, such as nonlinear (i.e., thresholding) properties for spike generation, integration of synaptic inputs at dendrites, and architecture of network connections, constrain the neural activity, therefore limit animals' capacity in recognizing the world. The goal of this article is to provide a unified Bayesian and thermodynamic view on these structured neuronal dynmamics involving organism's recognition and learning. We review key experimental observation of neural dynamics that support the Bayesian view of the learning and recognition in organisms, and explain that the neural dynamics that performs the Bayesian inference forms the recently-proposed information-theoretic engine (neural engine) \cite{shimazaki2015neurons,shimazaki2018neural}. We demonstrate that this new paradigm allows us to quantitatively assess animal's internal computation such as attention or awareness from neural spike data.} %

\en{It is thought that the brain is adapted to the characteristic statistical structure of natural stimuli to gain the information efficiently. This hypothesis, known as the efficient coding hypothesis \cite{barlow1961possible,barlow1972single}, suggests that the sensory systems in organisms attempt to convey the maximum amount of information of the external world by eliminating redundancy of external stimuli. The hypothesis is also known as the information maximization principle (Infomax principle) \cite{linsker1988self} that is the optimization principle of the nonlinear responses in nervous systems \cite{bell1995information}. For example, experiments with fly visual cells demonstrated that the nonlinear response functions of neurons adapted to the distribution of visual stimuli in the external world \cite{laughlin1981simple}. The hypothesis also also explains emergence of Gabor-like receptive fields of neurons in early visual systems \cite{linsker1988self,olshausen1996emergence,olshausen1997sparse}. It is this process of adapting nonlinear devices to the distributions of external signals that we call learning.}

\en{The natural environment is not stationary but continuously changing. Living organisms need to adapt to (or learn) the time-varying distributions of sensory inputs; therefore, they have mechanisms to adapt their nonlinear responses to the changing environment more quickly than the above-mentioned learning processes \cite{brenner2000adaptive}. The gain control, or gain modulation, is a ubiquitous adaptation mechanism that shifts or rescales the nonlinear response function under the new distribution. A typical example is gain control dealing with the intensity and contrast changes in visual stimuli. A visual stimulus has a dynamic range of 10 to the 9th power, ranging from $10^{-4}$ ${\rm cd / m^2}$ for the blink of a star at night to $10^{5}$ ${\rm cd / m^2}$ for sunlight during the day. Therefore, it is not easy to directly express the light intensity using the dynamic range of neurons from 0 Hz to several hundred heltz. Typically, the responses of neurons adapted to the dark environment saturate in the bright environment, resuling in low sensitivity to the bright light intensity. However, neurons in the retina and primary visual cortex can respond to a wide range of light intensities by adapting the nonlinear response function to the new stimulus distribution; phenomena called light adaptation and contrast gain control \cite{Sakmann1969,shapley1979contrast,shapley1979nonlinear,ohzawa1982contrast}. By this adaptation, the portion of the response function with the highest rate of change encodes the stimulus intensities/contrasts that most frequently appear. Thus, neural activity again becomes sensitive to changes in the stimulus. There are multiple mechanisms that realize light adaptation and contrast gain control, even in the retina alone.\footnote{These mechanisms include depletion of photopigment in photoreceptors and a shift from one type of photoreceptor (the rods) to the other (the cones). Also, the contrast gain control is thought to take place in both bipolar cells and retinal ganglion cells \cite{carandini2012normalization}.} Further, the pupillary light reflex of pupil diameters regulates the amount of light projected to the retina. Thus, organisms cope with changes in the external world through multiple adaptation mechanisms and acquire information from the external world as much as possible.}

\en{The gain control is a broad concept describing the nonlinear interaction of two or more inputs (signal sources) on a neuronal response \cite{silver2010neuronal}. It is the basis of information integration by the neural systems (nonlinear devices). As will be discussed in this article, the gain control also explains higher-order brain functions such as attention and coordinate transformation in visual systems \cite{salinas2001gain}. The ``divisive normalization'' that explains various characteristics of the sensory responses in cortical microcircuitries is also a part of the gain control. Therefore, not a single mechanism may be able to explain the variety of phenomena resulting in the gain control. Indeed, researchers proposed multiple mechanisms. Examples are the gain control realized by changes in the strength of background synaptic inputs \cite{chance2002gain,burkitt2003study}, shunting inhibition \cite{doiron2001subtractive,prescott2003gain,mitchell2003shunting}, and synaptic depression \cite{abbott1997synaptic,rothman2009synaptic}. Similarly to the gain control in engineering systems, neural systems can realize the gain control by either feedforward or feedback connections. The sophisticated cognitive functions achieved by the gain control are likely to be caused by the feedback inputs from higher cortical areas. The gain control by feedback modulation should involve dynamics with time-delay.}

\en{This article explains how the learning of nonlinear response functions extend to the process by which the brain acquires a statistical model of the external world; the view termed as the Bayesian brain hypothesis. It will be also shown that the Bayesian view of the stimulus-response provides a statistical paradigm for the gain control. According to the Bayesian brain hypothesis, the brain constructs a model of the external world (generative model) that assumes causes behind the sensory input, and makes inferences about the causes. The inference is achieved by constructing a posterior distribution of the causes using this model according to the Bayes' theorem. One can immediately see that the posterior distribution derived from the genrative model represents stochastic responseses of neurons with specific nonliear response functions (see Appendix of this article). The gain control is realized as changes of the generative model. Furthremore, similarly to the learning and gain control of the nonlinear response functions, the generative model is learned so that it acquires the distribution of external stimuli. In Shimazaki 2019 \cite{shimazaki2019principles}, we've seen that we could describe this adaptive process, once again, from the information-theoretic perspective as a process maximizing the amount of mutual information between the structure of the nervous system and the external world via neural activity.}

\en{When we consider the information-theoretic principles of the recognition and learning by the nervous systems of an organism, it is necessary to take into account their dynamics. Neurophysiological studies on early sensory cortices revealed that an initial feedforward-sweep of neural response depends on stimulus features whereas perceptual effect such as awareness and attention is represented as modulation of the later part of the stimulus-response. As introduced in this article, some of these experiments were described precisely in the framework of the gain control. The delayed modulation is presumably mediated by feedback connections from higher brain regions. We show that the delayed gain control of the stimulus response via recurrent feedback connections is modeled as a dynamic process of the Bayesian inference that combines the observation and top-down prior with time-delay.}

\begin{figure*}[ht]
\begin{center}

\en{ \includegraphics[width=.9\textwidth,pagebox=cropbox]{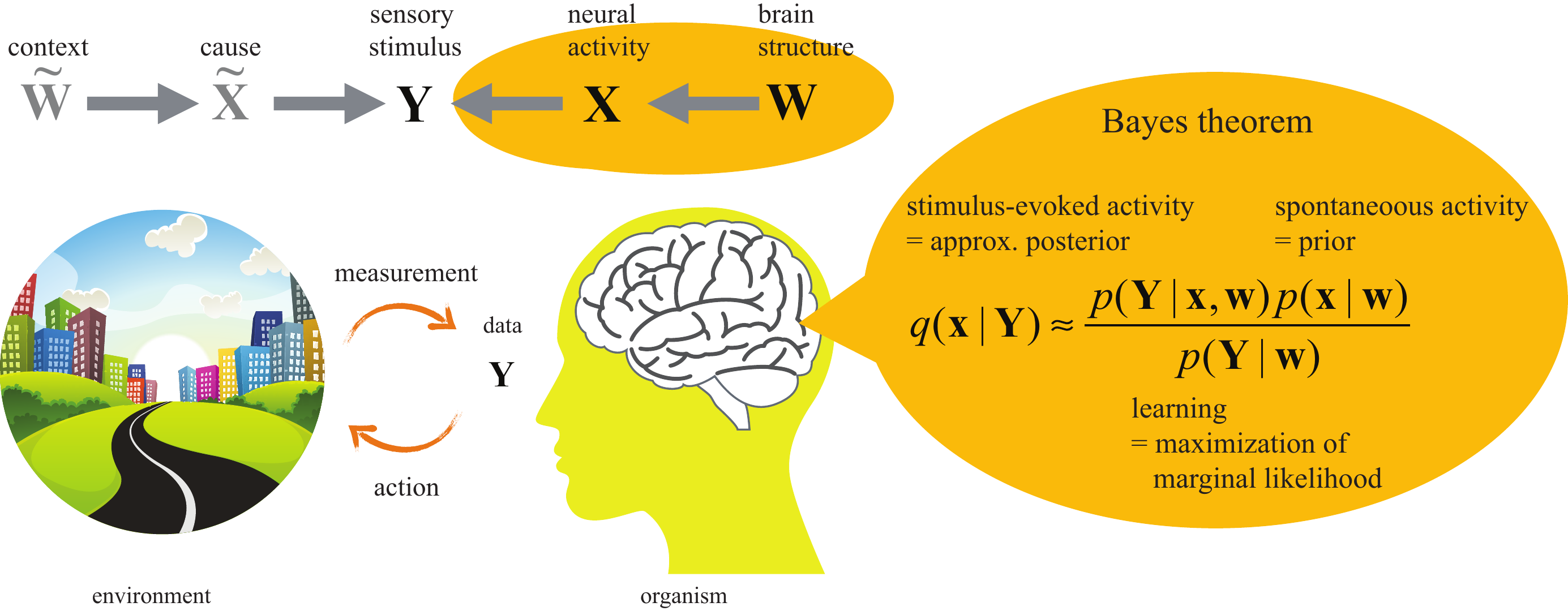} }
\end{center}
\caption{
 \en{A Bayesian view of neural activity}
}
\label{fig:bayesianbrain_evoke_spont_jp}
\end{figure*}

\en{Interestingly, it will be shown that this process becomes a mathematical analogue of a heat engine in thermodynamics. To see this, this article reviews a thermodynamic approach to the dynamics of learning and recognition. Here, the thermodynamic approach refers to describing learning and recognition by the laws about the entropy of neural activity, using a model obtained under the maximum entropy principle. Specifically, we derive the conservation law of entropy  (the first law) for the neural dynamics of recognition (stimulus-response) and show the conditions under which the law of increasing entropy (the second law) represents the learning. Based on the first law, we explain that the neural dynamics that realizes the Bayesian inference through feedback gain control behaves analogously to a heat engine. This view provides us to quantify the amount of the delayed gain control and its efficiency in retaining the stimulus information in terms of entropy changes of the neural activity. Thus it offers a way to measure the perceptual capacity of organisms.}

\en{The structure of this article is as follows. In Section \ref{sec:bayesianbrain}, we briefly introduce an interpretation of neural activity as the one that realizes the Bayesian inference. Next, we review experimental studies that support this view in Section \ref{sec:neural_learning_recognition}. We eplain how experimental studies lead to the idea that the brain constructs a generative model of the external world (Section \ref{sec:spontaneous_vs_evoked}), and investigate details of the stimulus-evoked neuronal dynamics by reviewing experiments that revealed dynamical integration of stimulus information with top-down prior knowledge (Section \ref{sec:inference_dynamics}). Finally, in Section \ref{sec:thermodynamics}, we construct the neural dynamics of learning and recognition using the statistical model of a spiking neural population. We introduce the thermodynamic approaches to these dynamics.}

\en{\section{A Bayesian view of the brain}\label{sec:bayesianbrain}}

\en{The Bayesian brain hypothesis states that the brain has a model of how causes generate external stimuli, and that the brain is an organ that makes inferences about the causes from an observed stimulus using this model. By assuming that neuronal activity represents the causes themselves as the most straightforward realization of this view, here we briefly introduce how nervous systems can realize the Bayesian inference (Fig.~\ref{fig:bayesianbrain_evoke_spont_jp}). See Shimazaki 2019 \cite{shimazaki2019principles} for more detailed descriptions of this section. We will explain how the Bayesian inference can be implemented by spiking (binary) neurons with nonlinear activation functions in Appendix and Section \ref{sec:thermodynamics}.}

\en{Let a set of the sensory stimuli (data) obtained from the outside world be $\mathbf{Y}_{1:n}=\{\mathbf{Y}_1,\mathbf{Y}_2,\ldots,\mathbf{Y}_n\}$, where $n$ is the number of samples. For simplicity, we use ${\mathbf{Y}}$ without the subscript to denote one of the samples. The sensory stimulus can be considered as a sample from the distribution $\bar p({\mathbf{y}})$. The organisms can access only the samples (i.e., empirical distribution of $\mathbf{Y}$), but can not directly access the underlying distribution $\bar p(\mathbf{y})$. We consider that the organism's brain is a device that builds a model of this unknown distribution based on past data and predicts the next sensory stimulus. The model that the brain constructs is denoted as $p({\mathbf{y}}|{\mathbf{w}})$, where ${\mathbf{w}}$ is a set of parameters representing the structure of the brain. The dynamics of learning is the process of changing the brain structure so that the model $p({\mathbf{y}}|{\mathbf{w}})$ approaches the \textit{true} distribution of the data $\bar p({\mathbf{y}})$. Namely, learning makes the distribution by the model closer to the true distribution. Here we use the Kullback-Leibler (KL) divergence as a criterion for the ``distance'' between distributions, and consider minimizing the KL divergence.\footnote{There is no evidence that supports the KL divergence is a metric suitable for natural data. In the future, one should extend the discussion in this article with more appropriate metric.} The minimization of the KL divergence is equivalent to maximizing the likelihood function with the observed data ${\mathbf{Y}_{1:n}}$ inserted into the model:}%
\begin{equation}
   {\mathbf{W}^{\ast}} = \arg {\max _{{\mathbf{w}}} } \log p({\mathbf{Y}_{1:n}}|{\mathbf{w}}). %
   \label{eq:maximum_likelihood_estimation}
\end{equation}
 \en{Under the assumption that the samples are independent, the log marginal likelihood above is computed as the sum of individual log marignal likehoods. We expect that, by the learning rule of the neural systems, the parameters approach the optimal values, ${\mathbf{W}^{\ast}}$.}

\en{The model distribution of sensory stimuli is constructed through neural activity. Consider $N$ neurons and denote the activity of the $i$th neuron by $x_i$.\footnote{We can think of $x_i$ as a continuous variable representing the firing rate of a neuron, or a discrete variable that takes a binary value ($0$ or $1$) or a count as spiking activity. Alternatively, one may consider a point-process model to represent the time of spike occurrence.} We denote the activity of the neural population by the vector ${\mathbf{x}}=[x_1,x_2,\ldots,x_N]'$. Consider the following hierarchical representation of a joint probability distribution (namely, a joint probability density/mass function) of sensory stimuli and neural activity,}
\begin{equation}
   p({\mathbf{y}},{\mathbf{x}}|{\mathbf{w}})=p({\mathbf{y}} \vert {\mathbf{x}}, {\mathbf{w}})p({\mathbf{x}} \vert {\mathbf{w}}). %
   \label{eq:generative_model1}
\end{equation}
\en{This joint distribution is called a generative model. $p({\mathbf{x}} \vert {\mathbf{w}})$ on the right-hand side represents neural activity that is not dependent on observation of the stimulus, which we call the prior distribution. Here, we assume that the spontaneous activity of neurons represents the prior distribution. $p({\mathbf{y}} \vert {\mathbf{x}}, {\mathbf{w}})$ is the representation of external stimulus by neural activity and is called an observation model. The model of the sensory stimulus $p({\mathbf{y}}|{\mathbf{w}})$ is obtained by integrating (marginalizing) the generative model with respect to the neural activity ${\mathbf{x}}$, and henceforth $p({\mathbf{Y}}|{\mathbf{w}})$ with the data assigned to the model is called the marginal likelihood function.}

\en{We consider the stimulus-evoked neural activity as a sample from the posterior distribution that represents a joint probability mass or density of the neural activity ${\mathbf{x}}$ given an observation ${\mathbf{Y}}$:}
\begin{equation}
   p\left( {\mathbf{x}}|{\mathbf{Y}},{\mathbf{w}} \right) = \frac{{p\left( {\mathbf{Y}}|{\mathbf{x}},{\mathbf{w}} \right)p\left( {\mathbf{x}}|{\mathbf{w}} \right)}}{{p\left( {\mathbf{Y}}|{\mathbf{w}} \right)}}.
   \label{eq:bayesian_formula}
\end{equation}
 \en{The posterior distribution depends on the parameter ${\mathbf{w}}$. Use of the parameter obtained by learning (Eq.~\ref{eq:maximum_likelihood_estimation}) allows appropriate inference (empirical Bayes' method). Eq.~\ref{eq:bayesian_formula} gives the exact posterior distribution. However, it is unlikely that the neural activity realized in the biological neural networks can constitute this exact posterior distribution. Rather, we consider that the neurons realize the one close to the exact posterior distribution for inference, and henceforth express the stimulus-evoked activity as an approximate posterior distribution $q\left( {\mathbf{x}}|{\mathbf{Y}} \right)$:}
\begin{equation}
   q\left( {\mathbf{x}}|{\mathbf{Y}} \right) \approx p\left( {\mathbf{x}}|{\mathbf{Y}},{\mathbf{w}} \right).
\end{equation}
 \en{The approximate posterior distribution $q\left( {\mathbf{x}}|{\mathbf{Y}} \right)$ is also called a recognition model.}

\en{We briefly introduced a Bayesian view of neural activity. The view presented here was not given top-down but gradually formed over the years, both from the theoretical development of the hierarchical models \cite{dayan1995helmholtz,olshausen1996emergence} and from experimental studies that revealed the relationship between spontaneous activity and stimulus-evoked activity. In the next section, we will introduce experimental studies that led to this view\footnote{Note that the hypothesis that the stimulus-evoked activity represents a posterior distribution is not the only hypothesis for the neural activity. For example, Rao \& Ballard's predictive coding theory \cite{rao1999predictive} likewise considers the neural mechanism that forms the posterior distribution based on the Bayes' theorem, but the neural activity itself represents the prediction error about the inferred causes rather than the causes themselves. The review by Aitchison \& Lengyel \cite{aitchison2017or} provides a summary of different statistical views on neuronal population activity.} and review the stimulus-evoked dynamics observed in neurophysiological experiments.}

\en{ \section{Experimental studies underpinning the Bayesian brain}\label{sec:neural_learning_recognition} }

\en{ \subsection{A brief research history of spontaneous and stimulus-evoked activities}\label{sec:spontaneous_vs_evoked} }

\begin{figure*}[!th]
\begin{center}

\en{ \includegraphics[width=1\textwidth,pagebox=artbox]{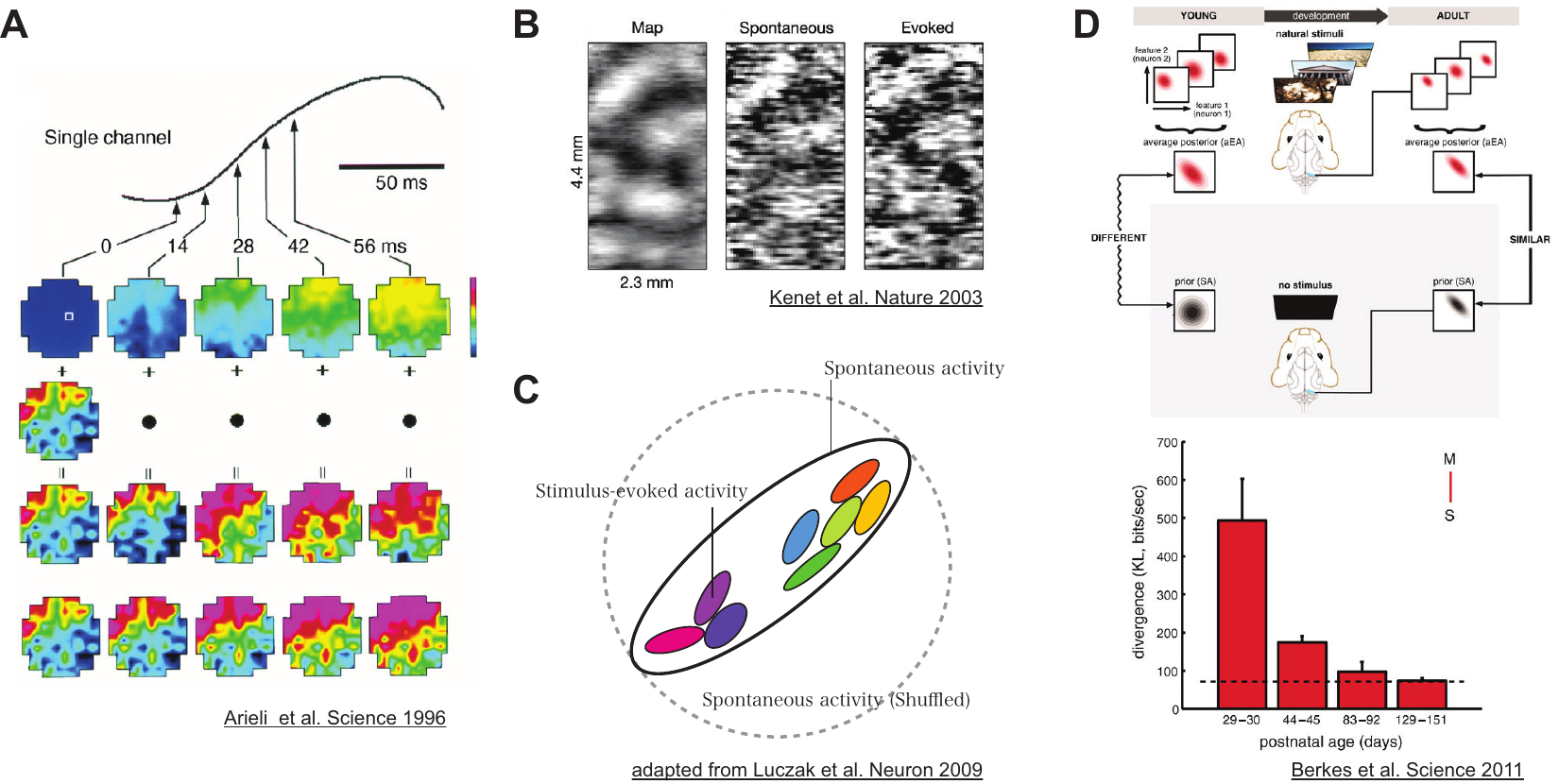} }
\end{center}
\caption{
 \en{\textbf{A} An experiment by Arieli et~al. examining the contribution of ongoing activity to stimulus-evoked activity in the early visual cortices (V1/V2) \protect\cite{arieli1996dynamics}. The curve above is trial average of stimulus-responses from a single channel of an optical imaging method. The maps in the lower four rows show neural activity measured by the optical recording method. The linear sum (third row) of the stimulus-dependent activity obtained as the trial average (first row) and the snapshot of ongoing activity immediately after the stimulus-response (initial state, second row) predicted the observed response to a stimulus (fourth row).}
 \en{\textbf{B} The similarity between spontaneous and stimulus-evoked activity discovered by Kenet et~al. \protect\cite{kenet2003spontaneously}. Left panel: An orientation map of the area V2 obtained by trial averaging. Central panel: A snapshot of spontaneous activity. Right panel: Stimulus-evoked activity. The orientation map was created by the stimulus orientation that induces the map most correlated with snapshots of the spontaneous activity. The stimulus-evoked activity was also obtained by this orientation.}
 \en{\textbf{C} A schematic of the relationship between possible spontaneous activities and stimulus-evoked activity found by Luczak et~al. \cite{luczak2009spontaneous}. The spontaneous activity here is the activity triggered by the occurrence of an UPSTATE. Because correlations constrain neural activity (black ellipse), the taken area by the spontaneous activity is narrower than that of uncorrelated activity with the same variance (dotted circle).  Individual stimulus-evoked activities (filled ellipses) have correlation structure (noise correlation) similar to the one in spontaneous activity. The stimulus-evoked activities fit into the realm of spontaneous activity.}
\en{\textbf{D}  Development of an optimal internal model revealed by Berkes et~al. \protect\cite{berkes2011spontaneous}. Top panel: The average of neural activities evoked by various natural stimuli (top left, red) are different from spontaneous activity in the dark (bottom left, black) in immature animals, but the two activities are expected to be similar in mature animals. Bottom panel: The KL-divergence between the average stimulus-evoked activity and spontaneous activity decreased with postnatal days.}
{\footnotesize %
\textbf{A} From Amos Arieli, Alexander Sterkin, Amiram Grinvald, and Ad Aertsen. Dynamics of on- going activity: explanation of the large vari- ability in evoked cortical responses. Science, 273(5283):1868-1871, 1996.. Reprinted with permission from AAAS.
\textbf{B} Reprinted by permission from Springer Nature: Nature Spontaneously emerging cortical representations of visual attributes, Tal Kenet et al, Copyright \copyright 2003, Springer Nature (2003)
\textbf{D} From Pietro Berkes, Gerg{\H{o}} Orb{\'a}n, M{\'a}t{\'e} Lengyel, J{\'o}zsef Fiser, Spontaneous Cortical Activity Reveals Hallmarks of an Optimal Internal Model of the Environment, Science, 331(6013):83-87, 2011. Reprinted with permission from AAAS.
}
}
\label{fig:spontaneous_vs_evoked}
\end{figure*}

\en{In a famous experiment published in 1996, Arieli et al. found that endogenous or internal dynamics explained a significant fraction of stimulus-evoked activity in the brain, and argued that these internal dynamics play an essential role in the brain's cognitive functions \cite{arieli1996dynamics}. In this experiment, they measured the visual cortex of anesthetized cats exposed to visual stimuli, using a combination of electrodes and an optical recording method with voltage-sensitive dyes. They showed that the stimulus-response separates into two components: internally-driven ongoing activity and stimulus-dependent activity (Fig.~\ref{fig:spontaneous_vs_evoked}\textbf{A}). The internal dynamics significantly contributed to the response dynamics, and the stimulus-dependent activity appeared on top of this internal activity. In this paper, Arieli et al. wrote that ``the effect of a stimulus might be likened to the additional ripples caused by tossing a stone into a wavy sea'', which presented a new way of viewing a stimulus response as a perturbation to the internal dynamics (see also \cite{fiser2004small}). This view was a departure from the traditional input-output view on the neural response to a stimulus, and is consistent with the Bayesian view of the brain, in that we see the stimulus-evoked activity is a sample from the posterior distribution. If we look at the Bayes' formula (Eq.~\ref{eq:bayesian_formula}), this equation can be seen as the process in which the spontaneous activity representing the prior distribution is modulated by observation, resulting in the stimulus-evoked activity that represents the posterior distribution. In this sense, their study already predicted the importance of the spontaneous activity on shaping recognition as a prior distribution of the Bayesian inference.}

\en{Arieli and colleauges presented these results in an attempt to answer the controversial question in the 1990s about cortical variability: Why do neurons show spiking activity that is highly variable and less reproducible even if we presnt the same stimulus? \cite{softky1993highly,shadlen1998variable}. However, this study initiated subsequent studies that further clarified the relationship between spontaneous activity and stimulus-evoked activity, which promoted the Bayesian view of the brain, as explained below. }

\en{Spontaneous activity has a certain statistical structure rather than random noise \cite{arieli1995coherent}. Then what is that structure? Through the series of experiments by Arieli and colleagues, it was becoming clear that the structure of spontaneous activity is similar to that of stimulus-evoked activity. For example, Tsodyks et al. showed that using the same experimental techniques as in the above-mentioned paper, the collective neural activity that activates single neurons has similar structure during spontaneous and stimulus-evoked activity in the visual cortex \cite{tsodyks1999linking}. More directly, Kenet et al. showed that the collective activity at a given moment while neurons fired spontaneously was very similar to the collective activity when a stimulus was presented (Fig.~\ref{fig:spontaneous_vs_evoked}\textbf{B}) \cite{kenet2003spontaneously}. Neurons in the early visual cortices shows selectivity to stimulus orientations. Therefore, the optical recordings reveal the orientation map of the recorded area. In this experiment, Kenet et al. showed that, in a statistically significant manner, the spontaneous activity exhibits activities resembling the stimulus-responses (orientation maps) of neurons to grating stimuli with specific orientations. The authors argued that the spontaneous activity with the characteristic structure, such as those representing the orientation selectivity, shape the internal state (context) of the brain.}

\en{Rather than merely describing the similarity between snapshots of the spontaneous and evoked activities, a study by Luczak et al. represented the population activity as a point in a high-dimensional probability space and examined the relationship of the distribution of the points obtained from stimulus-evoked activity with the distribution obtained from spontaneous activity. More precisely, Luczak et~al. examined the relationship between the spontaneous and stimulus-evoked activities of the rat auditory cortex in response to auditory stimuli (natural stimuli and artificial tones) using the simultaneous recordings of 45 neurons. They found that the responses to auditory stimuli fit in the space that the spontaneous activity could take place (Fig.~\ref{fig:spontaneous_vs_evoked}\textbf{C}). This result makes sense if we consider that the stimulus-evoked activity represents a conditional probability that is obtained by conditioning a prior distribution representing the spontaneous activity by the stimulus.}

\en{In the experiments of Luczak et~al, they used the limited number of stimuli. If we extrapolate their observations by assuming that we could use more stimuli, then it may be expected that the sum of the stimulus-evoked activities fills the state space that spontaneous activity can take place. In math, it is expressed as:}
\begin{equation}
\int q\left( {\mathbf{x}}|{\mathbf{y}}\right) \bar p\left( {\mathbf{y}} \right) d{\mathbf{y}} \approx p\left( {\mathbf{x}}|{\mathbf{w}} \right).
\label{eq:evokedavg_vs_spontaneous}
\end{equation}
 \en{Two conditions are required for an exact equality to hold in the above equation. First, the stimulus-evoked activity of neurons (recognition model) must represent an exact posterior distribution, i.e., the inference must be exact: $q\left( {\mathbf{x}}|{\mathbf{Y}} \right)=p\left( {\mathbf{x}}|{\mathbf{Y}},{\mathbf{w}} \right)$. Second, the model distribution of sensory stimuli (data) assumed from the generative model must represent the distribution of data in the external world: $p\left( {\mathbf{Y}}|{\mathbf{w}} \right) = \bar p\left( {\mathbf{Y}} \right)$. In this case, Eq.~\ref{eq:evokedavg_vs_spontaneous} is the exact formula that the prior distribution is obtained by marginalizing the generative model with respect to the stimuli:\footnote{However, this paper identifies the stimulus-evoked activity with the exact posterior distribution and considers only construction of an optimal generative model by learning.}}
\begin{equation*}
\int p\left( \mathbf{x}|\mathbf{y},\mathbf{w} \right) p\left( \mathbf{y}|\mathbf{w} \right) d\mathbf{y} = p\left( \mathbf{x}|\mathbf{w} \right).
\end{equation*}
 \en{Thus, close distributions at the left and right sides of Eq.~\ref{eq:evokedavg_vs_spontaneous} suggests that the organism possesses an optimal internal model and performs optimal inference. Could such a thing really be happening?}

\en{By an experiment with ferrets, Berkes et al. reported that the average of the stimulus-evoked activities in response to various stimuli and the spontaneous activity indeed approached to each other as the animals develop (Fig.~\ref{fig:spontaneous_vs_evoked}\textbf{D}) \cite{berkes2011spontaneous}. In this experiment, they recorded neural activity in the visual cortex when presenting movies to awake ferrets under free-viewing conditions. They also recorded spontaneous activity in the dark. The recordings were made at the time of eye-opening (postnatal days 29-30), completion of orientation selectivity (P44-45), and two postmaturity periods (P83-92 and P129-151). The results confirmed that the KL divergence between the average stimulus-evoked activity (left side of Eq.~\ref{eq:evokedavg_vs_spontaneous}) and spontaneous activity (right side of Eq.~\ref{eq:evokedavg_vs_spontaneous}) decreased with development.}

\en{Based on these and other experimental results, the view has been formed that sees the spontaneous activity as sampling from a prior distribution and the stimulus-evoked activity as sampling from a posterior distribution, and that the optimal internal model is constructed through learning.\footnote{However, see also \cite{stringer2019spontaneous} for the recent large-scale analysis on stimulus-evoked and ongoing activity of more than $10,000$ neurons in mouse visual cortex, which provided an opposing result.}}

\en{ \subsection{Recognition dynamics}\label{sec:inference_dynamics} }

\en{If we assume that the brain realizes the Bayesian inference by expressing the posterior distribution using stimulus-evoked activity, it is necessary to consider the neural dynamics leading to such posterior distribution. This is because the stimulus-evoked activity of neurons must follow the dynamics constrained by neuronal mechanisms. Emphasis on the dynamics is in contrast with some of the machine learning approaches in which researchers can obtain the approximate posterior analytically.\footnote{However, some other sampling methods such as the Markov chain Monte Carlo method follow dynamics; therefore, these algorithms can potentially be implemented in neural systems.} In this section, we present experimental results suggesting that the brain dynamicaly integrates information about an input stimulus with prior knowledge. These experiments show that this process of integrating information is closely related to animals' cognitive capacities. In the next section, we will introduce a hierachical Bayesian model that describes the neural dynamics presented here.\footnote{Note that here we introduce studies based on the classical experimental paradigm in which the effects of the stimulus-independent internal dynamics are removed by trial averaging. However, we will take into account the internal (or recurrent) dynamics that is triggered by stimuli and tasks.}}

\en{Two types of inputs constitute the dynamics of neural activity. One is the input from feedforward connections, which are projected sequentially from the sensory organs that govenrn the organism's senses, to the lower and then to the higher brain regions. The other type of input comes from recurrent connections, which include horizontal and lateral connections within the same column or cortical area, and feedback connections from the higher areas. The inputs from the feedforward and recurrent connections should arrive at the target region with different timings.}

\en{In the case of the visual system, the initial response to a high-contrast visual stimulus reaches the final stage of the ventral pathway within about 100 ms \cite{lamme2000distinct,thorpe2001seeking}.\footnote{The response latency of visual neurons depends on the stimulus intensity. However, it is highly variable among individual neurons \cite{gawne1996latency,reynolds2000attention,lee2007spatial}. In addition, the definition and detection methods significantly affect estimated values of the latency. Here, we present the values from the cited paper. The results of meta-analyses of the various literature can be found in Lamme \& Roelfsema \cite{lamme2000distinct}. Finally, the input to the MT field via the superior colliculus and the pulvinar, which is the pathway different from the ventral pathway, takes less than 40 ms. So the visual information processing takes place with multiple time-scales.} It is thought that the feedforward inputs cause the initial stimulus response. In the visual areas (e.g., V1, V2, and V4), the difference in response latencies between the lower cortical area and the next area is about 10 ms, but the feedback input from the higher region to the lower area takes about 10 ms, so the effect of the feedback may not catch up to the initial resposne \cite{lamme2000distinct}. Based on this fact, we can examine the computations performed primarily by the feedforward networks from properties of neuron's initial responses. It turns out that the magnitude of the initial response depends only on the characteristics and intensity of the stimulus and not related to the cognitive functions of the animal (see below). For example, it was suggested that the orientation selectivity of the classical receptive field of monkey V1 neurons is formed without inputs from the feedback connections from higher cortical areas \cite{celebrini1993dynamics}.\footnote{Note that mechanisms for the formation of orientation selectivity by visual neurons differ among species significantly. For example, the orientation selectivity is formed in the visual cortex in monkeys and cats. However, in mouse, rabbit, and zebrafish, the orientation selectivity is already found in the retina \cite{scholl2013emergence,antinucci2018orientation,johnston2019retinal}.} The computations that can be performed by the feedforward connections should not be underestimated. Indeed, the remarkable performance of artificial convolutional neural networks (CNNs) in recognition tasks, which imitate operations of only feedforward and horizontal connections, has demonstrated the high computational power of a cascade of nonlinear units without feedback connections \cite{yamins2014performance}.}

\en{Nevertheless, it has been pointed out that recursive inputs through feedback connections play an essential role in various cognitive functions of the brain. The influence of inputs from the feedback connections should appear after the feedforward input caused the initial response. However, the dynamics after the initial response are not only influenced by the feedback connections but also by other neuronal mechanisms. For example, spiking mechanisms of an individual neuron may undergo sensitization resulting in reduction of the firing rate. If this happends at presynaptic neurons, then the feedforward inputs may be already attenuated. In order to dissociate these confouding mechanisms, researchers constructed thoughtful experiments that can identify the effect of the delayed inputs that are presumably mediated by feedback circuits. In the following, we present studies that identified the impact of feedback inputs on the stimulus-response and its relationship to animal's cognitive capacities.}

\begin{figure}[!t]
\begin{center}
\includegraphics[width=.5\textwidth]{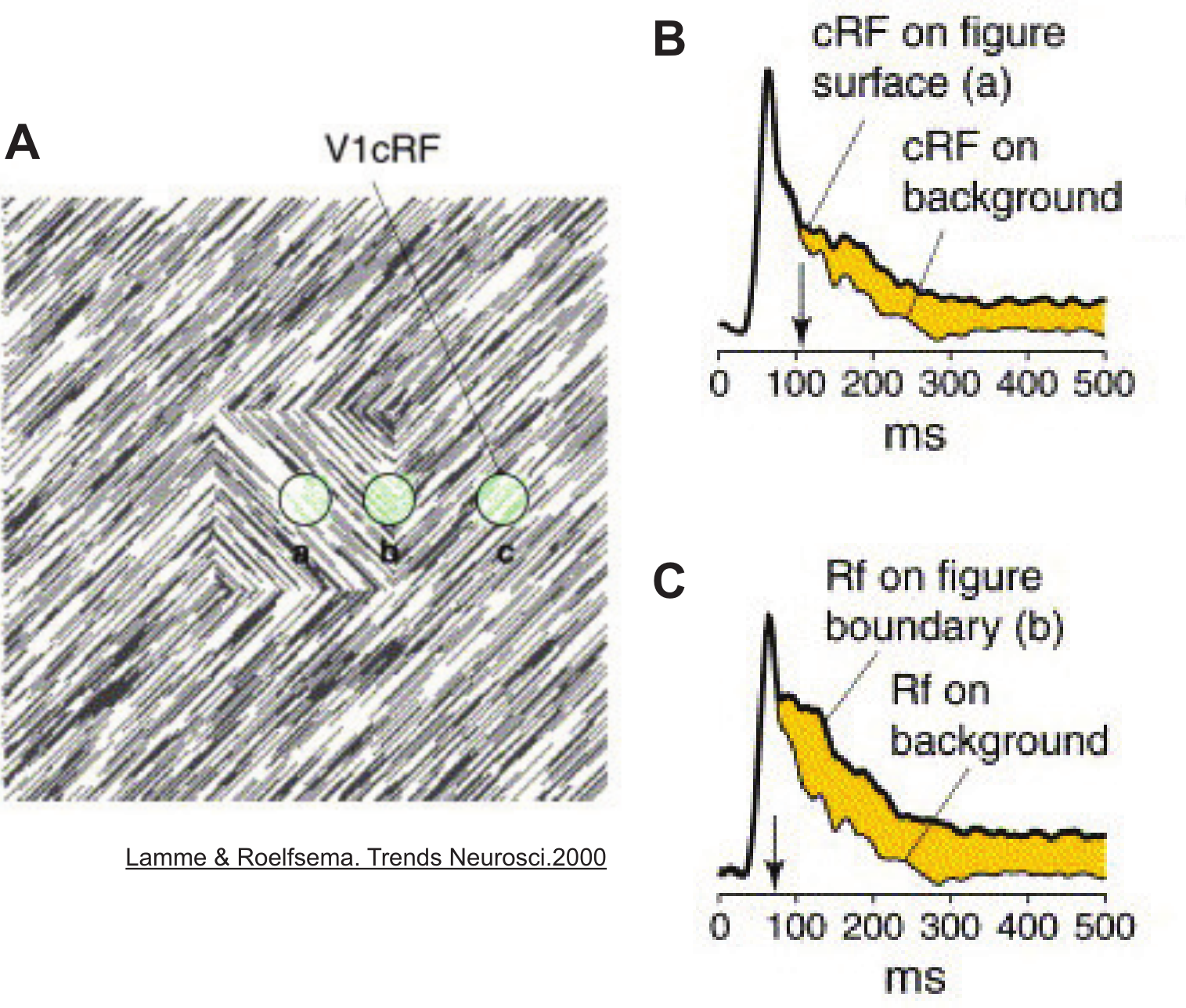}
\end{center}
\caption{ \en{
Delayed modulation of the stimulus-response of V1 neurons as revealed by the Lamme et al.'s figure-ground separation experiments \protect\cite{lamme1999separate,lamme2000distinct}
\textbf{A} Location of the receptive fields of V1 neurons in relation to the stimulus used for the figure-ground separation. A figure is shown in this example. The figure disappears when the oriented lines in the background exhibit the same direction as those at the center. Recordings were made when the receptive field was (a) in the figure, (b) on the border, and (c) in the background.
\textbf{B} Comparison of the stimulus-responses (average of 32 electrode recordings) of neurons whose receptive fields are in the figure, while the orientations of the figure and ground are different (surface) or same (background). Differences in response appeared from approximately 100 ms \protect\cite{lamme2000distinct} (112.5-122.5 ms \protect\cite{lamme1999separate}).
\textbf{C} Comparison of the stimulus-responses when the receptive field is at the boundary of the figure, while the orientation of the figure and the ground are different (surface) or they match (background). The responses differed after approximately 80 ms \protect\cite{lamme2000distinct} (90 ms \protect\cite{lamme1999separate}).}
{\footnotesize 
Reprinted from Trends in neurosciences, 23(11), Victor AF Lamme and Pieter R Roelfsema, The distinct modes of vision offered by feedforward and recurrent processing, 571-579, Copyright (2000), with permission from Elsevier.
}
}
\label{fig:lamme00}
\end{figure}

\en{\textbf{Perception of visual stimuli} A series of experiments by Lamme and his colleagues suggested that perception of a visual stimulus requires integration of the stimulus inputs by feedforward connections with delayed inputs by recurrent connections conveying contextual information \cite{lamme2000distinct} (Fig.~\ref{fig:lamme00}). Lamme et al. used a visual stimulus paradigm called figure-ground separation \cite{lamme1995neurophysiology} to examine stimulus responses in the primary visual cortex (V1) of awake monkeys. They found that the context of the visual stimulus outside the classical receptive fields of V1 neurons was expressed in the response activity 80-100 ms after the onset of the visual stimulus \cite{lamme1995neurophysiology}. The presence or absence of the modulation in this part of the response correlated with reports on the perceptual experience (awareness) of the stimulus by the monkeys \cite{zipser1996contextual,super2001two}.}

\en{As a visual stimulus, on the one condition, Lamme and colleagues presented segmented lines with a specific slope (figure) on top of a background image (ground) that is also composed of segmented lines but perpendicular to the figure (Fig.~\ref{fig:lamme00}\textbf{A}). They presented the ``figure'' within the classical receptive fields of the recorded V1 neurons. On the other condition, they tilted the ground so that the orientaion of the segmented lines in the ground becomes the same as of the figure. In this case, the distinction between the figure and ground disappears, although the stimuli within the classical receptive field are the same under the two conditions. When they compared the responses of neurons to the stimulus with or without the separation of the figure and ground, they found no difference in the firing rates in its early component starting from about 30-40 ms, but the firing rates of late components starting from 80-100 ms differ: The `figure' stimulus induced the larger activity at the later part of the stimulus-response (Fig.~\ref{fig:lamme00}\textbf{B}) \cite{zipser1996contextual,lamme1999separate,super2001two}.\footnote{A similar effect was observed not only in the orientation but also in the figure-ground separation using disparity, color, and luminance as a context \cite{zipser1996contextual}.} In this article, we call this later part of the response a ``late component'' and change of the late component due to conditions on the presented stimulus, ``delayed modulation''. A closer look at the temporal evolution of the delayed modulation reveals that the modulation starts at 80 ms after the stimulus onset if the classical receptive field is at the boundary of the figure and 100 ms after the stimulus onset if it is in the figure \cite{lamme1999separate,poort2012role} (Fig.~\ref{fig:lamme00}\textbf{A}). These modulations occur later than 50 ms, at which neurons show selectivity to the orientation. From these resutls, Lamme and colleagues suggested that the information processing takes place in the order of orientation selection, boundary detection, and figure detection.}

\en{Feedback input from higher cortical areas is thought to cause this delayed modulation rather than connections within V1/V2 because lesion of the extrastriate cortex beyond V2 eliminated modulation of the response after about 100 ms (late component) involved in figure detection \cite{lamme1998feedforward}.\footnote{But, see also \cite{macknik2004dichoptic} that showed the late components of early monocular visual neurons are not related to the feedback inputs by masking experiments in which the target and mask are presented to different eyes. The authors proposed that late component is created as the after-discharge caused by lateral inhibitory networks without feedback inputs \cite{macknik2004spatial}. Involvement of feedback inputs in the late component may depend on the type of stimulus and how complex the context is.} Since the initial sweep in the entire ventral pathway completes within about 100 ms, the late component of the response in V1 after 100 ms can reflect computations done in the inferior temporal cortex and beyond, which makes it more likely that the late component reflects perceptual experiences. The following experimental results support the hypothesis that, not the initial response, but this delayed modulation is indeed involved in perceptual experiences. 
(i) In experiments presenting different figure-ground stimuli to two eyes (binocularly rivalrous displays), the delayed modulation depends on the perceived figure (or ground) caused by its combination, and not on the response expected for the figure-ground stimulus presented to the one eye \cite{zipser1996contextual}. 
(ii) Even if animals are exposed to the stimulus in which the figure and ground are separated, the delayed modulation after 100 ms selectively disappears in neural responses of trials in which the animals failed to report the presence of the figure \cite{super2001two}. 
(iii) The larger the amount of the delayed modulation (at responses involving figure detection) is, the shorter the reaction time of the saccade used to report the figure, and the higher the accuracy of the report \cite{poort2012role}. 
(iv) In monkey experiments using backward masking with pattern masks, the accuracy of the figure detection correlated with magnitude of the delayed modulation reduced by the masks \cite{lamme2002masking}. 
(v) Anesthetization selectively removed the late component related to the figure detection. The initial response (including responses involving boundary detection) and neurons' receptive field were unaffected by the anesthesia \cite{lamme1998figure}.\footnote{Experiments using rats reported that the initial response ($0-100$ ms) of neurons in the visual cortex remains unchanged as the concentration of anesthesia (desflurane) increases, but the late component ($150-1000$ ms) is selectively lost under anesthesia \cite{hudetz2009desflurane}.}
}

\en{In sum, their results suggest that the late component in the responses of V1 neurons 100 ms after the stimulus onset was affected by feedback inputs from the extrastriatal and inferior temporal cortex, which may reflect the global context involving the perceptual experience. To the contrary, neurons in the superior temporal sulcus of the temporal lobe display information on global features such as the face of a person or a monkey 91 ms after the stimulus onset, and 51 ms later they integrate detailed information about the individuals and facial expression. These results suggest that, after 100 ms of stimulus onset, the entire recurrent circuit of visual areas, including the inferior temporal cortex and beyond, is used to perform computations for recognition.}

\en{\textbf{Perception of tactile stimuli} The perceptual experience of a tactile stimulus also involves modulation of the late component of the neuronal response in somatosensory systems \cite{cauller1995layer}. Moreover, by selectively eliminating the late component using optogenetic stimulation, it was found that presence of the late component causes perceptual experience. Sachidhanandam et al. reported in vivo recordings of the barrel cortex in the mouse primary somatosensory cortex (S1) and found that the membrane potential responding to a whisker stimulus had an initial response component less than $50$ ms and a delayed response component of $50$-$400$ ms. The presence or absence of the late component in the membrane potential correlated with whether the mice reported the sensory stimuli correctly or not (HIT or MISS). This was also true for the late component of the postsynaptic firing rates. However, no such relationship was found for the initial component of the response. Furthermore, when they selectively suppressed the late component 100 ms after the stimulus onset by optogenetic stimulation, the animals reduced the perceptual reports, confirming that this late component is one of the causes of the perceptual experience \cite{sachidhanandam2013membrane}.}

\en{Manita et al. identified the circuit underpinning the late component and demonstrated that the feedback input from a higher cortical area is necessary for reporting the perceptual experience \cite{manita2015top}. Manita et al. recorded the response of S1 neurons to electrical stimulation of hind legs of mice. They found that the response was clearly separated into an initial response that appeared about 20 ms after the stimulus onset and a late component that occurred at about 110 ms later. Using pharmacological and optogenetic methods, they demonstrated that the late component is caused by a reverberation circuit that projects from S1 to the secondary motor area (M2) and returns to S1. Manita et al. also found a causal effect of the late component on sensory recognition. Inhibition of bilateral projections from M2 to S1 under free moving conditions eliminated the mice's inherent preference for a certain texture of the tactile stimuli and worsened their performance on a tactile stimulus discrimination task. In addition, when they selectively suppressed the late component by optogenetic stimulation, the mice did not respond to electrical stimulation to the hind legs. These results indicate that the initial response is insufficient for tactile perception, and the delayed input by the recurrent circuit from M2 to S1 is required for accurate recognition of the tactile stimuli.\footnote{The response of mouse S1 to electrical stimuli completely separated into the initial and late components, therefore does not integrate information from the stimulus with recurrent inputs. However, the late component may act as a primer for the subsequent stimulus. Similarly, if the feedback input from M1 to S1 is unspecified, it may also act as priming for stimuli to other sites.}}

\begin{figure*}[!t]
\begin{center}
\includegraphics[width=.9\textwidth]{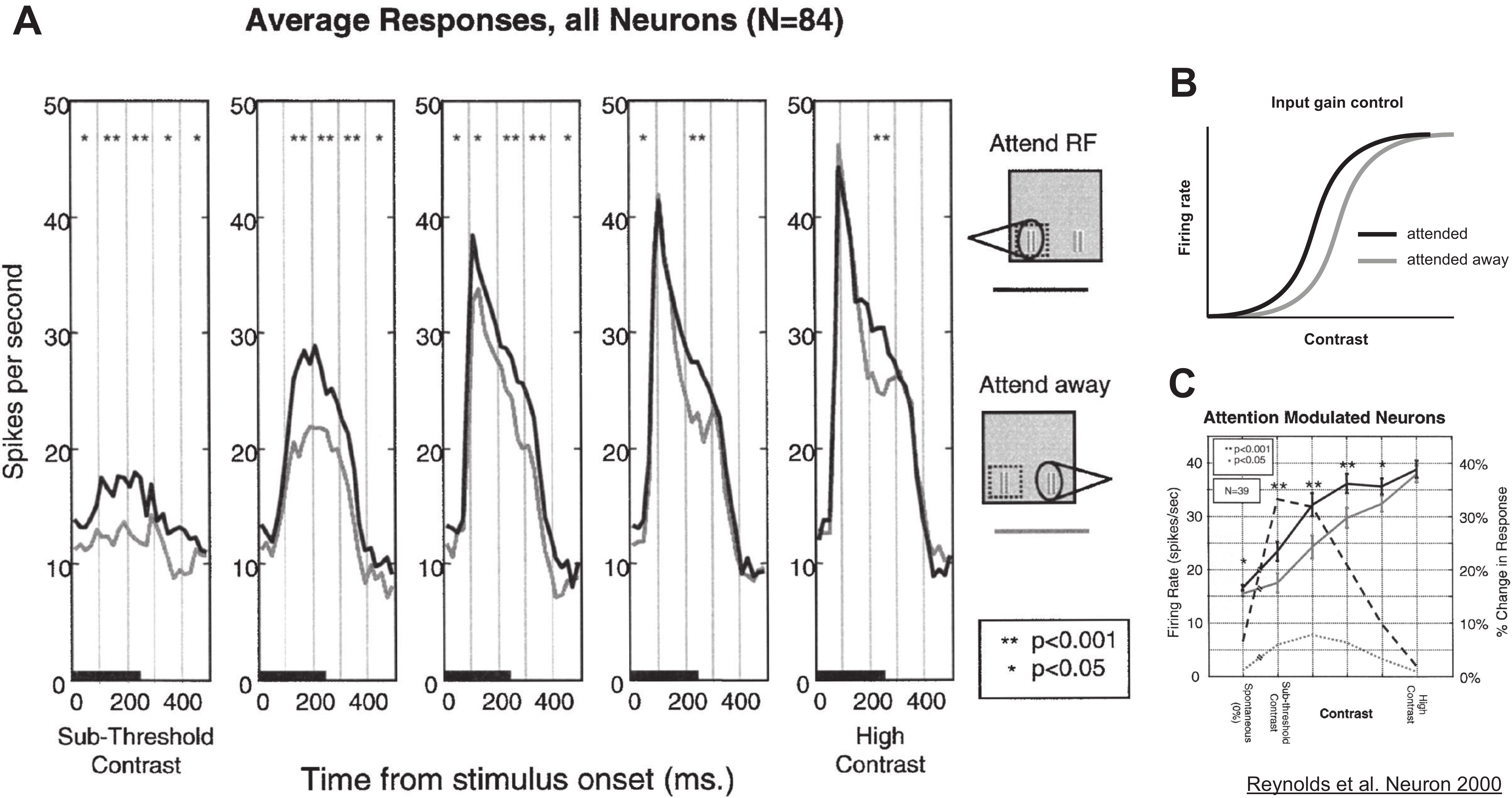}
\end{center}
\caption{ \en{
An experiment by Reynolds et al. examining the effect of attention on the stimulus response of monkey V4 neurons. 
\textbf{A} Average stimulus-response from 84 neurons. The solid black line is the response when monkeys directed attention to the receptive field of recorded neurons. The solid gray line is the response when attention is directed away from the receptive field. The panels show, from left to right, the response of neurons to, from low to high, contrast stimuli. The stimuli were presented for 250 ms. Asterisks at the top of the panels are the test results on the difference between responses with and without attention by dividing the observation period into 100 ms intervals.
\textbf{B} A schematic diagram of the gain control by attention. 
\textbf{C}} Average firing rate up to 400 ms after the onset of stimulus presentation. The solid black and gray lines are the firing rates with or without attention, respectively. It displays the average of 39 neurons that showed significant changes by attention. Reproduced with permission from \protect\cite{reynolds2000attention} Fig.4, 5.
{\footnotesize 
Reprinted from Neuron, 26(3), John H Reynolds, Tatiana Pasternak, and Robert Desimone, Attention increases sensitivity of V4 neurons, 703-714, Copyright (2000), with permission from Elsevier.
}
}
\label{fig:reynolds00}
\end{figure*}

\en{\textbf{Top-down attention} It is known that, with attention, animals can actively modulate the late component. Attention is a neural mechanism that facilitates processing of specific stimuli. In particular, top-down attention refers to the mechanism by which animals can actively select an object to be processed \cite{maunsell2015neuronal}. In the case of visual attention, there are three types of attentions: spatial attention directed at a part of the space, feature-based attention directed at a feature of the stimulus such as color or shape, and object-based attention. The effect of top-down visual attention is more pronounced in higher visual areas, and in general the attention elevates responses of neurons with receptive fields to which attention is directed \cite{moran1985selective}. Here, we first introduce an experiment by Reynolds et al. that examined the effect of spatial attention on the response of monkey V4 neurons when a single stimulus is presented within the classical receptive field.} 

\en{In Reynolds et al.'s experiment, they presented two simultaneous stimuli to monkeys fixating at the center of the screen (Fig.~\ref{fig:reynolds00}\textbf{A}, see schematics on the right). At the start of a trial, a cue stimulus is presented to an animal that gives an indication as to whether to pay attention to the left or right stimulus. They recorded the activity of neurons containing one of the presented stimuli in their receptive field. Each panel in Fig.~\ref{fig:reynolds00}\textbf{A} shows dynamics of the neuronal responses under different stimulus contrasts (average of 84 neuronal responses). The black line represents the response when attention is directed to the receptive field, and the gray line shows the response when attention is directed outside the receptive field. For example, at the highest contrast, we find no difference in the responses up to 200 ms after the stimulus onset between the conditions with and without attention. However, we observe an effect of attention in the period from 200 ms to 300 ms after the stimulus onset.}

\en{The response of neurons to a stimulus increases with strength of the stimulus contrast, which is represented as a nonlinear response function of the neurons (Fig.~\ref{fig:reynolds00}\textbf{B} gray line). When the monkey directs attention to the neurons' receptive field, it alters the response function because the firing rate at each contrast increases. The resulting shift of the response function to the left caused by attention is called contrast gain control (Fig.~\ref{fig:reynolds00}\textbf{B} black line). Another possibility is response gain control, in which the firing rate increases even at high contrasts without saturation. Reynolds and colleuagues confirmed that the contrast gain control explains the effect of attention on the response function, rather than the response gain control  (Fig.~\ref{fig:reynolds00}\textbf{C}).\footnote{In this experiment, they used a suboptimal stimulus; therefore, the firing rates of neurons are not saturated even at the highest contrast. That is, this mechanism is not a simple gain control caused by the response saturation. Also note that other studis suggest that the effect of attention is represented by the response gain control \cite{maunsell2015neuronal}.} They also showed that the attention increased the statistical power of neurons to detect stimuli.\footnote{The task was to detect appearance of a tilted grating stimulus among vertical grating stimuli. The sensitivity (estimation accuracy) to continuous stimulus features depends on the magnitude of the first-order derivative (slope) of the response function with respect to the feature. Attention reduced the sensitivity to contrast changes in the medium to high contrast, where we observe the large effect of attention. Therefore, the increased firing rate due to attention may be used to increase the sensitivity of other stimulus features than the contrast. For example, for neurons with Gaussian (von Mises) orientation selectivity, increasing the firing rate in a multiplicative manner increases the orientation sensitivity (at non-optimal orientation). We can not argue the effect of attention on the information coding of the stimulus unless we specify the neuronal response function to the stimulus feature and how the attention changes the function.} In conclusion, Reynolds et al. showed that the higher cognitive function such as attention is explained by the gain control, which is the basis of adaptation mechanisms of neural systems. Further, this gain effect appears later in the response activity. Therefore the dynamics of attention is represented by the \textbf{delayed gain control}.}

\en{Many experiments show that other types of attention than the spatial attention can also modulate the late component. Motter's study was the first to show that attention can be directed to stimulus features and reported that the attentional effect appears in the sustained firing activiy during stimulus presentation \cite{motter1994neural}. In this experiment, monkeys performed a task to detect the slope of rod-shaped visual stimuli with different colors while recording the activity of V4 neurons. The firing rate of neurons 150-200 ms after the stimulus onset increased when the characteristics of the stimulus in the receptive field were consistent with those to which monkeys directed attention. In this experiment, they found no effect of attention during the initial response that appeared 50 ms after the stimulus onset nor duing a preparation period.} %

\begin{figure}[!t]
\begin{center}
\includegraphics[width=.4\textwidth]{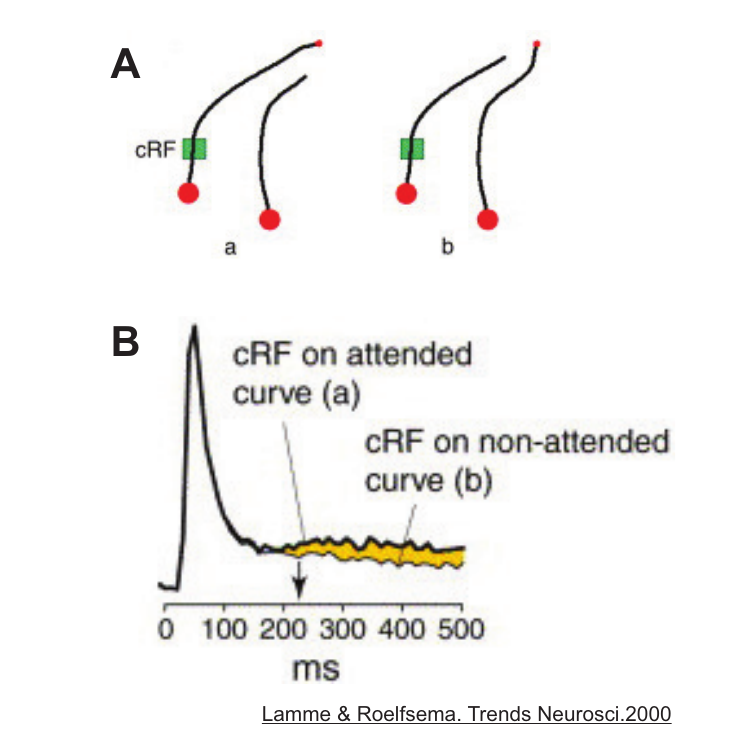}
\end{center}
\caption{ \en{
An experiment by Roelfsema et al. to investigate the effect of object-based attention \protect\cite{roelfsema1998object,lamme2000distinct}.
\textbf{A} A monkey gazes at the fixation point (a small red circle) and moves its gaze to a point (a large red circle) at the end of the curve (target) connected to the fixation point 600 ms after the onset of stimulus presentation. Left panel: The classical receptive field of V1 neurons shown in the square is on a curve connected to the fixation point. Right panel: The receptive field is on the curve that is not connected to the fixation point.
\textbf{B} An average stimulus-response of V1 neurons (mean of the normalized responses from multiple recordings). Even though the stimuli were identical within the receptive field, the activity rates differ when the receptive field was on the attended object (the curve connected to the solid viewpoint) and when the receptive field was on an unattended object. This difference did not show up in the initial response, but appeared 235 ms after the stimulus onset. Note that the normalization was calculated by using the peak values and spontaneous firing rates on the average of the responses of two stimulus conditions. If there is a difference in the maximum firing rates between stimuli, the normalized peak value will also show a difference.}
{\footnotesize 
Reprinted from Trends in neurosciences, 23(11), Victor AF Lamme and Pieter R Roelfsema, The distinct modes of vision offered by feedforward and recurrent processing, 571-579, Copyright (2000), with permission from Elsevier.
}
}
\label{fig:roelfsema1998object}
\end{figure}

\en{The classic example of the object-based attention is an experiment by Roelfsema and colleagues examining the responses of monkey V1 neurons in a curve-tracing task \cite{roelfsema1998object}. In this experiment, they presented two curves to monkeys: One is a curve connected to a fixation point and the other is a curve not connected to the fixation point (Fig.~\ref{fig:roelfsema1998object}\textbf{A}). The end points are marked by red circles. After 600 ms has passed since the onset of stimulus presentation, the monkeys are trained to saccade the eyes to an endpoint of the curve  connected to the fixation point (target). During the stimulus presentation, one of the two curves enters the receptive field of V1 neurons (green sqaures). Note that the stimuli in the receptive fields were identical under the two conditions. Comparing responses of neurons with and without the receptive field on the target showed no difference in the initial responses starting at 35 ms after the stimulus onset. However, the firing rate increased after 235 ms if the receptive field of a neuron is on the target (Fig.~\ref{fig:roelfsema1998object}\textbf{B}). When the stimulus curves were arranged to intersect each other, the delayed modulation was still observed for neurons with the receptive field on the target, suggesting that V1 neurons participate in the computation for recognizing smoothly connected lines as the same object in the late part of the response.}

\en{Poort et al. investigated the effects of attention on neuronal activity related to the aforementioned figure-ground separation. This study showed that attention increases the V1 neurons' response involved in detection of the figures 204 ms after the stimulus onset, whereas the attention modulated activity of V4 neurons at 159 ms after the stimulus onset. That is, the effect of attention was seen earlier in V4 than in V1 \cite{poort2012role}.\footnote{V1 neurons show an initial response at 40 ms after the stimulus onset, delayed modulation to detect a boundary at 60 ms, and delayed modulation to detect a figure at 95 ms. These are followed by modulation by attention at 204 ms. In V4, the initial response was observed at 52 ms after the stimulus onset, followed by delayed modulation at 67 ms, and modulation by attention at 159 ms} From this result, Poort et al. proposed a model in which figure detection occurred at V1 from 95 ms, and the effect of attention at V4 that appears at 159 ms reached V1 and enhances figure detection from 204 ms.}

\en{These experimental results show that attention causes time-delayed modulation of the responses of V1 and V4 neurons. Most examples of the delayed modulations presented here reported no change in spontaneous activity before stimulus presentation as well as in the intial rising phase of the response. Hence the observed delayed modulation is likely to be caused by feedback input driven by the stimulus. However, a substantial number of literatures also report that, even before presenting the stimulus, attention significantly elevates spontaneous activity of neurons whose receptive fields contain the attended features \cite{luck1997neural,lee2007spatial,maunsell2015neuronal}. The presence of the attentional effect on the spontaneous activity indicates that the presentation of a cue stimulus that initiates the task has a prolonged effect that extends across trials \cite{luck1997neural}. These dynamics, too, can be seen as delayed modulation, which, however, requires the dynamics with more extended time-scale used for task understanding and memory retention.}

\en{\textbf{Reward values} Finally, several studies postulate that reward value is also represented by the late component. Schultz proposes that the response activity of dopamine neurons consists of an initial and late components, and that the late component represents subjective reward value and utility \cite{schultz2016dopamine,schultz2017phasic}. The 40-120 ms early response component of dopamine neurons represents stimulus intensity, while the 150-250 ms late component represents value \cite{fiorillo2013multiphasic}. Under challenging tasks, the delayed response appears much later and completely separated from the initial response, indicating computing the value takes a longer period of time \cite{nomoto2010temporally}. Schultz proposes that such a two-component response allows animals to respond quickly to rewarding stimuli while eliminating the possibility of error as much as possible.}

\en{ \section{Thermodynamics of the Bayesian brain}\label{sec:thermodynamics} }

\en{So far, we looked over experimental results suggesting that organisms acquire a generative model of the external world by learning (Section \ref{sec:spontaneous_vs_evoked}), and that the dynamics of feedforward and recurrent inputs realizes integration of input information with internal states (Section \ref{sec:inference_dynamics}). In this section, we explain these experimental results as dynamics that realizes the Bayesian inference by introducing a generative model constructed from spiking activity of neurons.}

\en{In particular, bearing in mind the attention experiment by Reynolds et al. \cite{reynolds2000attention}, we reproduce the delayed gain control by attention as the dynamics of the Bayesian inference. We then show that this dynamics constitutes an information-theoretic engine, termed `neural engine', which works analogously to a heat engine in thermodynamics \cite{shimazaki2015neurons,shimazaki2018neural,shimazaki2019principles}. Based on this paradigm, we explain how to quantify internally-defined quantities such as awareness, attention, and reward value from neural data. Finally, we provide an unified thermodynamic view on the learning and inference, deriving laws about the entropy of neural activity by reconstructing the process of integrating a stimulus input and prior knowledge under the maximum entropy principle.}

\en{ \subsection{The Bayesian inference by a neural population}\label{sec:multiple_neurons} }

\en{We describe the spike response of a population of neurons using a generative model. See Appendix for the relationship between the response function, gain control, and the generative model using a simpler, independent neurons. Below we consider the behaviour of a population of correlated neurons. Consider $N$ neurons. We represent the state of the $i$th neuron by ${x_i} = \{ 0,1\}$, and the activity pattern of the population by ${\mathbf{x}} = [{x_1},{x_2}, \ldots ,{x_N}]'$. The input stimulus is represented by the $d$-dimensional column vector ${\mathbf{Y}}$. The generative model is composed of an observation model and a prior. We construct the observation model that represents the input stimulus ${\mathbf{Y}}$ by the combination of the neurons, using a normal distribution:}
\begin{equation}
p({\mathbf{Y}}|{\mathbf{x}},  \boldsymbol{\Sigma}, {\boldsymbol{\Phi}} ) = \frac{1}{\sqrt{|2\pi \boldsymbol{\Sigma}|}}{e^{ - \frac{1}{2}({\mathbf{Y}} - {\boldsymbol{\Phi}}{\mathbf{x}})'\boldsymbol{\Sigma}^{-1}({\mathbf{Y}} - {\boldsymbol{\Phi}} {\mathbf{x}})}}.
\label{eq:observation_ising}
\end{equation}
 \en{We used $\boldsymbol{\Sigma}$ as the covariance matrix. ${\boldsymbol{\Phi}}$ is a $d \times N$ matrix with an array of the basis vector for each neuron (${\boldsymbol{\Phi}} =[\boldsymbol{\phi}_1, \boldsymbol{\phi}_2, \ldots, \boldsymbol{\phi}_N]$). ${\boldsymbol{\Phi}} {\mathbf{x}}$ is a representation of a stimulus by the population of neurons. Next, we represent the prior distribution of neuronal activity by the Ising model,\footnote{The Ising model is a model that represents the distribution of binary patterns. The distribution contains no statistical structure other than those specified by the first and second-order statistics. Such a distribution belongs to the maximum entropy model. The Ising model is a standard model introduced in statistical physics as a model for interacting magnetic spins. It later became a model for consolidation and retrieval of memory, and used in machine learning today.}}
\begin{equation}
p({\mathbf{x}}|\boldsymbol{\Omega} ) = {e^{\frac{1}{2}{\mathbf{x'}}\boldsymbol{\Omega} {\mathbf{x}} - \psi_0 (\boldsymbol{\Omega} )}}.
\label{eq:prior_ising}
\end{equation}
 \en{Here $\boldsymbol{\Omega}$ is a matrix of $N \times N$ dimension, where the diagonal and off-diagonal components are the bias and second-order interaction terms, respectively. The posterior distribution based on the generative model consisting of Eqs.~\ref{eq:observation_ising} and \ref{eq:prior_ising} is written as follows:}
\begin{align}
&p({\mathbf{x}}|{\mathbf{Y}},{\mathbf{w}}) 
= \frac{{p({\mathbf{x}}|\boldsymbol{\Omega} )p({\mathbf{Y}}|{\mathbf{x}},\boldsymbol{\Sigma}, \boldsymbol{\Phi})}}{{p({\mathbf{Y}}| \mathbf{w} )}} \nonumber\\
&\propto \exp \left[ {\frac{1}{2}{\mathbf{x'}}\boldsymbol{\Omega} {\mathbf{x}} + ({\mathbf{Y}}' {\boldsymbol{\Phi}}\boldsymbol{\Sigma}^{-1}{\mathbf{x}} - \frac{1}{2}{\mathbf{x'}} {\boldsymbol{\Phi}}' \boldsymbol{\Sigma}^{-1} {\boldsymbol{\Phi}} {\mathbf{x}}) } \right]. 
\label{eq:exact_posterior_ising}
\end{align}
 \en{In this equation, we did not show terms that do not depend on ${\mathbf{x}}$. We defined a set of the model parameters as $\mathbf{w}=\{\boldsymbol{\Omega},\boldsymbol{\Sigma},\boldsymbol{\Phi}\}$. This posterior distribution becomes once again an Ising model.}

\en{Neural activity sampled from this posterior distribution achieves a correct inference for the external cause ${\mathbf{x}}$. However, in the above derivation, we did not consider the dynamics leading to this exact posterior distribution. Here, to consider the dynamics forming the posterior distribution given by Eq.~\ref{eq:exact_posterior_ising}, we extend the generative model as follows. In the following, we fix the parameter ${\mathbf{w}}$.}

\en{First, we replace the covariance matrix of the observation model $\boldsymbol{\Sigma}$ with $\alpha^{-1} \boldsymbol{\Sigma}$, where $\alpha$ is a scaler parameter. For example, when $\boldsymbol{\Sigma}=\mathbf{I}$ ($\mathbf{I}$ is a unit matrix), $\alpha$ is a parameter for the accuracy. Next, we replace the parameter $\boldsymbol{\Omega}$ in the prior distribution with $\beta \boldsymbol{\Omega}$ using a scalar $\beta$. By these extensions, we can describe the dynamics of recognition leading to the optimal inference by the dynamics of $\beta$ and $\alpha$. Let the posterior distribution based on this new generative model be the recognition model representing neural activity. We can express it as:}
\begin{align}
q&({\mathbf{x}}|{\mathbf{Y}})
= \frac{{p({\mathbf{x}}|\beta \boldsymbol{\Omega} ) p({\mathbf{Y}}|{\mathbf{x}}, \alpha \boldsymbol{\Sigma}, \boldsymbol{\Phi} )}}{{p({\mathbf{Y}}|{\mathbf{w}}, \beta, \alpha )}} \nonumber\\
&\propto \exp \left[ {\frac{\beta}{2}{\mathbf{x'}}\boldsymbol{\Omega} {\mathbf{x}} + \alpha ({\mathbf{Y}}' {\boldsymbol{\Phi}} \boldsymbol{\Sigma}^{-1} {\mathbf{x}} - \frac{1}{2}{\mathbf{x'}} {\boldsymbol{\Phi}}' \boldsymbol{\Sigma}^{-1}  {\boldsymbol{\Phi}} {\mathbf{x}}) } \right]. 
\label{eq:approx_posterior_ising}
\end{align}
 \en{A posterior distribution based on the original generative model (Eq.~\ref{eq:exact_posterior_ising}) is constituted when $\beta=1$ and $\alpha=1$. The posterior distribution is identical to the prior distribution when $\beta=1$ and $\alpha=0$, namely when the variance of the observation model diverges. Thus, one could realize the dynamics leading to the original posterior distribution (Eq.~\ref{eq:exact_posterior_ising}) from spontaneous activity after the stimulus onset by changing $\alpha$ from $0$ to $1$ under $\beta=1$. Without loss of genrality, we can consider that the orignal posterior distribution offers the optimal inference.\footnote{Given $\mathbf{w}$, it is possible to search the optimal $\beta$ and $\alpha$ for each sample under the marginal likelihood principle. One can alway rescale $\boldsymbol{\Sigma}$ and $\boldsymbol{\Omega}$ to make $\beta=1$ and $\alpha=1$ at the optimal distribution.} Thus this is a process in which the recognition model dynamically approaches the posterior distribution that realizes the optimal inference. Here we assume that neural activity in each state follows an equilibrium distribution (Eq.~\ref{eq:recognition_exp}). This process is called a quasi-static process.\footnote{One can think of several approaches to realizing the dynamics of neural activity that approaches the optimal posterior distribution. As an analytical method, one can consider the gradient dynamics leading to the MAP estimate of the posterior distribution. Many of the inferences based on the predictive coding theory of Rao \& Ballard and the free energy principle of Friston et al. belong to this type of inference. Here, as a complementary to this idea, we consider that the scaling parameters of the recognition model vary dynamically to change the distribution from a prior distribution (spontaneous activity) to a posterior distribution (evoked activity). The other approach is to use the sampling method, which samples neural activity from the targeted posterior distribution using the Markov chain Monte Carlo method, starting from an initial spontaneous state of a neural population. This process is a non-equilibrium process that depends on the spiking history. Both approaches construct approximate posterior distributions (recognition models).}}

\en{In the following, the recognition model (Eq.~\ref{eq:approx_posterior_ising}) is concisely wirtten as}
\begin{equation}
q({\mathbf{x}}|{\mathbf{Y}}) = \frac{1}{{{{Z}_{\beta ,\alpha }}({\mathbf{Y}})}}{e^{ -\beta  \mathcal{H}_0({\mathbf{x}}) + \alpha \mathcal{H}_1({\mathbf{x}})  }}, 
\label{eq:recognition_ising}
\end{equation}
 \en{where $\mathcal{H}_0({\mathbf{x}})$ is a feature derived from the prior distribution and $\mathcal{H}_1({\mathbf{x}})$ is a feature derived from the observation model:}
\begin{align}
\mathcal{H}_0({\mathbf{x}}) &=  -\frac{1}{2}{\mathbf{x'}}\boldsymbol{\Omega} {\mathbf{x}}, \\
\mathcal{H}_1({\mathbf{x}}) &=  {\mathbf{Y}}'\boldsymbol{\Phi} \boldsymbol{\Sigma}^{-1} {\mathbf{x}} - \frac{1}{2}{\mathbf{x'}}\boldsymbol{\Phi}' \boldsymbol{\Sigma}^{-1} \boldsymbol{\Phi} {\mathbf{x}}.
\end{align}
 \en{The expectation of a feature by the recognition model is called the expectation parameter. We define the expectation parameter as $\left< \mathcal{H}_0(\mathbf{x}) \right>= U$ and $\left< \mathcal{H}_1(\mathbf{x}) \right>= V$. Note that $\langle \cdot \rangle$ represents expectation by the recognition model. From now on, $U$ is called internal activity, and $V$ is called stimulus-related activity. Given that the set of parameters $\mathbf{w}$ is fixed, the recognition model becomes a function of $\beta$ and $\alpha$. That is, we can represent the dynamics of this neural population in a two-dimensional space. One can also control the distribution by the expectation parameters $U$ and $V$, instead of $\beta$ and $\alpha$.}

\begin{figure}[!t]
\begin{center}

\en{ \includegraphics[width=.5\textwidth]{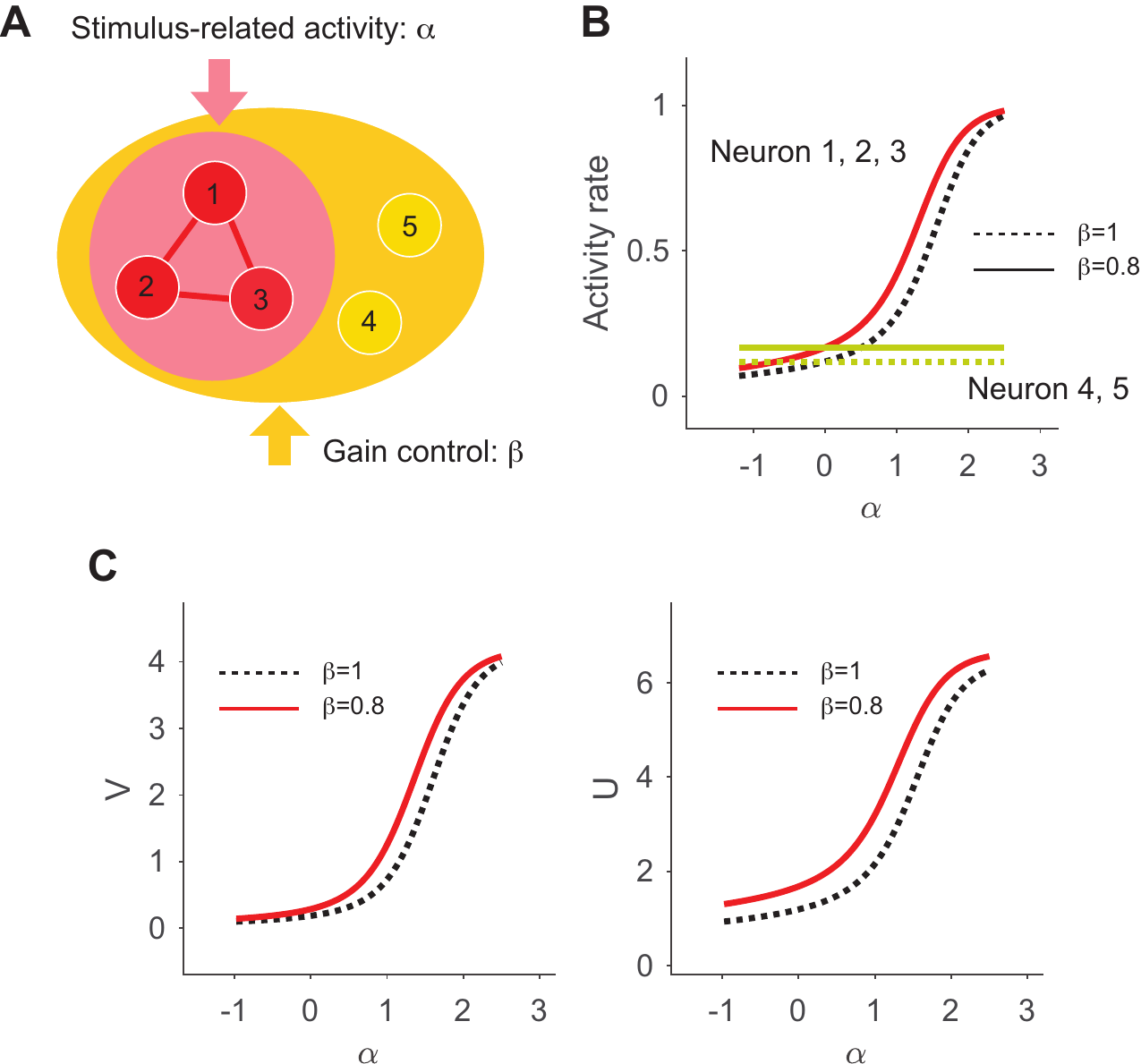} }
\end{center}
\caption{ \en{The gain control by a model of neural population. 
   \textbf{A} A group of five neurons. Neurons 1, 2, and 3 in red change their activity rates and correlations according to $\alpha$. The activity rates of all neurons, including yellow neurons, is controlled by $\beta$. 
   \textbf{B} The dependency of the activity rates of neurons 1, 2, 3, or neurons 4, 5 on $\alpha$. The dotted line is obtained at $\beta=1$. The solid line is $\beta=0.8$, which increases the overall activity level. The change in $\beta$ realizes the gain control for neurons 1, 2, 3. 
   \textbf{C} The relationship between the expectation parameters $V$ and $U$ and $\alpha$. The expectation parameters are also subject to the gain control by $\beta$. 
   We used a diagonal matrix with the diagonal component $[-2,-2,-2,0,0]$ as the parameter $\boldsymbol{\Omega}$ for the prior distribution. The parameters of the observation model were ${\mathbf{Y}}=0.1$, $\boldsymbol{\Sigma}=\mathbf{I}$, and $\boldsymbol{\Phi}=[1,1,1,0,0]$. Created using the parameters given by the genearative model with reference to Fig.~1 in \protect\cite{shimazaki2015neurons,shimazaki2018neural}.}
}
\label{fig:ising_example}
\end{figure}

\en{As an example, we explain the behavior of the recognition model (Eq.~\ref{eq:recognition_ising}) by considering a population of five neurons (Fig.~\ref{fig:ising_example}\textbf{A}). We define the parameters of the generative model such that the activity features related to the internal activity ($\mathcal{H}_0({\mathbf{x}})$), and the feature for the stimulus-related activity $\mathcal{H}_1({\mathbf{x}})$ are given as follows. $\mathcal{H}_0({\mathbf{x}})$ contains the bias term for all neurons and does not include the interactions. We note that the smaller $\beta$ is, the higher the activity rates are. Next, $\mathcal{H}_1({\mathbf{x}})$ contains bias and interaction terms related to neurons 1, 2, and 3, which means that increasing $\alpha$ above 0 increases activity rates of neurons 1, 2, and 3 as well as their correlations. In other words, it induces a cell assembly encoding the stimulus.}

\en{The dotted line in Fig.~\ref{fig:ising_example}\textbf{B} shows the activity rate of each neuron $\langle x_i \rangle$ $(i=1,2,\ldots,5)$ when  we change $\alpha$ while fixing $\beta$ at 1. The activity rates of neurons 1, 2, and 3 increase with $\alpha$ and saturate at the upper limit of 1. To the contrary, since the stimulus-related activity does not include neurons 4 and 5, they are constant regardless of $\alpha$. The solid line represents the activity rates of neurons when we increase the overall activity level by setting $\beta=0.8$. By decreasing $\beta$, the activity rates of neurons 4 and 5 increase. Similarly, the activity rates of neurons 1, 2, and 3 increase, but the response function shifts to the left because the activity rates are subject to the upper bound. That is, we realize the gain control of the response function by adjusting the activity rates of all neurons using $\beta$.\footnote{The function shown in Fig.~\ref{fig:ising_example}\textbf{B} is the response function with respect to $\alpha$. Similarly to this, we can realize the gain control of the response function with respect to the input stimulus $\mathbf{Y}$ by controling $\beta$. In Reynolds et al.'s experiment \cite{reynolds2000attention}, they observed the input gain control even though the response did not saturate at high contrast because they used non-optimal stimuli. Therefore, we need to set the maximum activity rate for each orientation to reproduce this phenomenon more precisely.} Fig.~\ref{fig:ising_example}\textbf{C} shows the internal activity $U$ and stimulus-related activity $V$ when we change $\alpha$ under $\beta=1$ or $\beta=0.8$. We observe similar gain control with respect to these average features.}

\en{ \subsection{Neural Engine: Recognition dynamics with delayed gain modulation}\label{sec:neuralengine} }

\begin{figure*}[!t]
\begin{center}

\en{ \includegraphics[width=.9\textwidth]{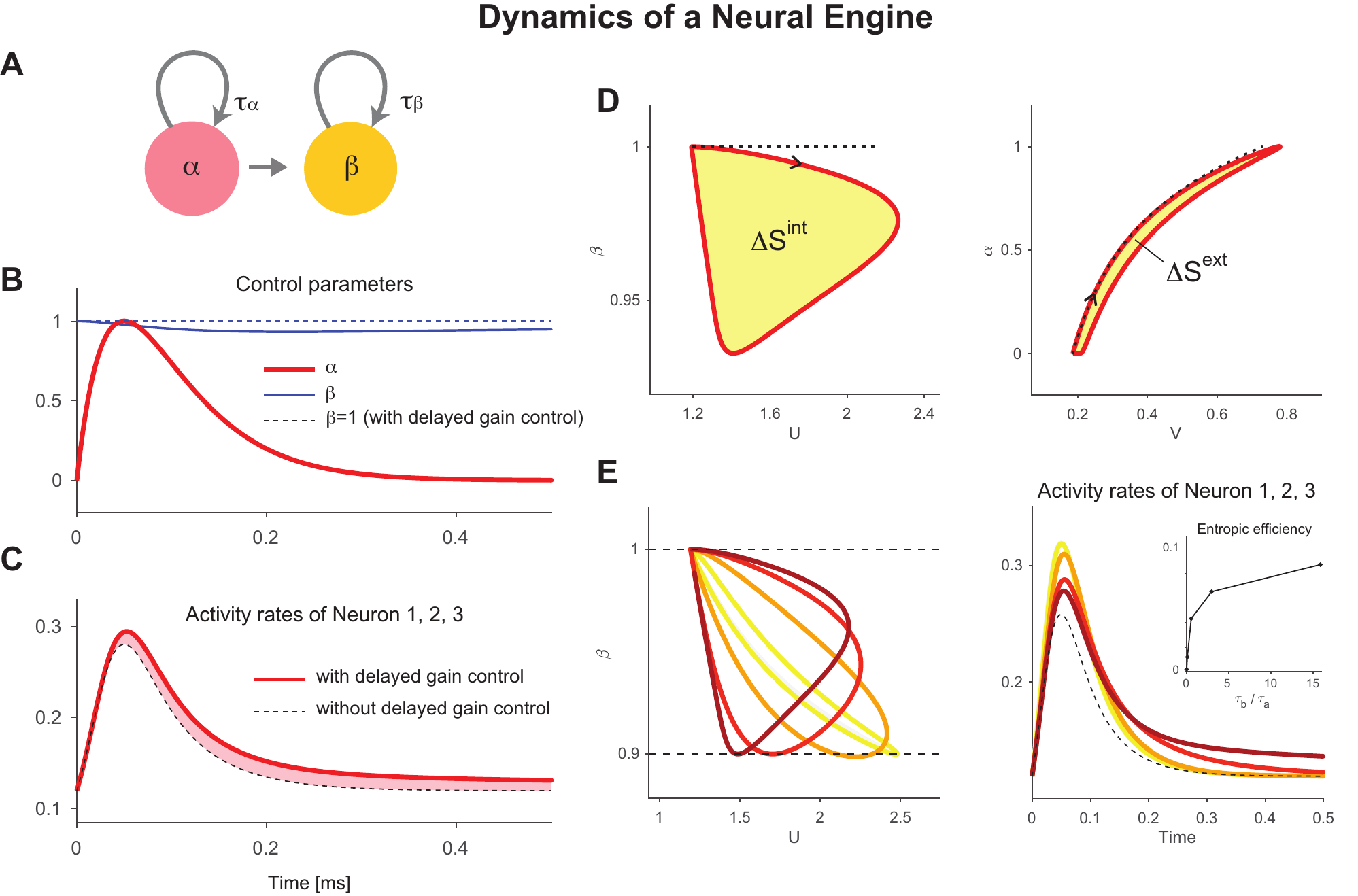} }
\end{center}
\caption{ \en{The information-theoretic engine realized by neural activity (neural engine).  The number of neurons and parameters of the features are the same as in Fig.~\ref{fig:ising_example}.
\textbf{A} A schematic diagram of the dynamics of the parameter $\alpha$ that regulates the stimulus-related activity, and the parameter $\beta$ that regulates the spontaneous activity.
\textbf{B} Dynamics of $\alpha$ and $\beta$. ($\alpha$ : thick solid line, $\beta$ : thin solid line and dotted line)
\textbf{C} Activity rates of neurons 1, 2, and 3. The dotted line shows the case without the gain control, and the solid line shows the case with the gain control with time-delay. The filled area is their difference.
\textbf{D} Phase diagram of the dynamics of the stimulus-response. The dotted line is without gain control. The cycle shown by the solid line is with the delayed gain control. Left panel: $U$-$\beta$ phase diagram. Right panel: $V$-$\alpha$ phase diagram.
\textbf{E} Left panel: $U$-$\beta$ phase diagram drawn for dynamics in which $\beta$ has a different time constant $\tau_{\beta}$. The minimum value of $\beta$ is fixed at $0.9$. Right panel: The activity rates of neurons 1, 2, and 3 for different time constants, $\tau_{\beta}$. The dotted line is the activity rate without the gain control. (inset) Entropic efficiency at each time constant. The dotted line shows the maximum efficiency $0.1$ realized with $\beta=0.9$. Created using the parameters given by the genearative model with reference to \protect\cite{shimazaki2015neurons,shimazaki2018neural} Fig.~2.}
}
\label{fig:delayed_gain}
\end{figure*}

\en{Using the statistical stimulus-response model of a neural population defined in the previous section, we now construct neural dynamics with time-delayed modulation as a dynamic process of the Bayesian inference, using time-varying $\beta$ and $\alpha$. The dyamics reproduces the stimulus-response to which the gain control is applied with time-delay as observed in the experiment with attention tasks \cite{reynolds2000attention}. We then explain how this cognitive dynamics constitutes an information-theoretic engine \cite{shimazaki2015neurons,shimazaki2018neural}. For this purpose, we first introduce the conservation law for entropy of neural activity.}

\en{Let the entropy of stimulus-response activity of a neural popluation be}
\begin{equation}
S=- \sum_{{\mathbf{x}}} q({\mathbf{x}}|{\mathbf{Y}})  \log q({\mathbf{x}}|{\mathbf{Y}}).
\label{eq:entropy_recognition}
\end{equation}
 \en{Entropy is dictated by the expected parameters $U$ and $V$ as natural independent parameters. Its total derivative can be written as follows (see the next section and references \cite{shimazaki2015neurons,shimazaki2018neural,shimazaki2019principles} for the derivation),}
\begin{align}
dS %
&= \beta dU - \alpha dV.
\label{eq:first_law_dS_0}
\end{align}
\en{This equation is a conservation law for entropy, and explains how much the total entropy of a neural population changes by changing its internal or stimulus-related activities. Eq.~\ref{eq:first_law_dS_0} posseses a mathematical structure similar to the first law of thermodynamics, the conservation law of energy. In thermodynamics, the temperature $T=1/\beta$ is introduced as a parameter to regulate the whole system by introducing $\alpha=\beta f$, where $f$ is known as a force applied to the system. The conservation law is then written as $TdS=dU-fdV$. The reason for represnting the conservation in this form in thermodynamics is that the system (gas) is controlled by the heat $\dbar Q=TdS$ and the work $\dbar W=fdV$. Here we do not introduce the concepts of heat and work because we think that the neural dynamics is controlled by $\beta$ and $\alpha$ (or the entropy of neural activity is controlled by $\dbar S^{\rm int} =\beta dU$ and $\dbar S^{\rm ext} =\alpha dV$). Nevertheless, we shall keep in mind that they share the same mathematical structure.\footnote{It should be noted that, by introducing $\alpha=\beta f$, we can use $\beta$ and $f$  as independent variables instead of $\beta$ and $\alpha$. With these paramters, we can achieve the gain control by changing $\beta$ in which only the sensitivity (slope) is controlled without shifting the response function horizontally  (see Appendix). In such a model of gain control, the concept of (inverse) temperature is valid because it regulates the stochasticity of the neural response (see next section).} The reason why they share the same mathmatical strucutre is because these distributions are derived from the maximum entropy principle (see next section). Following the first law of thermodynamics, we call Eq.~\ref{eq:first_law_dS_0} \textbf{the first law of neurodynamics}.}

\en{We now reproduce the cognitive dynamics with the delayed modulation introduced in Section \ref{sec:inference_dynamics} using the neural population defined in Section \ref{sec:multiple_neurons}. To this end, we introduce simple dynamics for the parameter $\alpha$, which regulates the stimulus-related feature, and the parameter $\beta$, which regulates the internal features (Fig.~\ref{fig:delayed_gain}\textbf{A}). For the dynamics of $\alpha$, in order to imitate the stimulus-response, we adopted dynamics that rapidly increases in the begening, followed by a decay in longer time-scale (Fig.~\ref{fig:delayed_gain}\textbf{B}).\footnote{We use the following dynamics in Fig.~\ref{fig:delayed_gain}\textbf{B} \cite{shimazaki2015neurons,shimazaki2018neural}
\begin{align}
& \tau_{\alpha}^2  \dot \alpha(t) = - \tau_{\alpha} \alpha(t) + s \, e^{-t/\tau_\alpha}, \nonumber\\ %
& \tau_{\beta} \dot \beta(t)   = - \beta(t) + \beta_0 -  \gamma \alpha(t), \nonumber  %
\end{align}
where the parameters are set as $\tau_{\alpha}=0.05$, $\tau_{\beta}=0.8$, $s=2.7$, $\gamma_0=1$, $\gamma=0.5$.}}

\en{The parameter $\beta$ responds slowly to this $\alpha$. Namely, the time constant $\tau_\beta$ for the dynamics of $\beta$ is larger than the time constant $\tau_\alpha$ for the dynamics of $\alpha$. Since the dynamics of the gain control is slow, the initial response realizes $\beta \approx 1$ and $\alpha \approx 1$, at which the stimulus-response realizes nearly optimal inference. Fig.~\ref{fig:delayed_gain}\textbf{C} represents the activity rates of neurons 1, 2, and 3. The solid line indicates the activity rate with the delayed gain control, and the dotted line is without the delayed gain control. Note that for the parameters used in the simulation, the smaller $\beta$ is, the larger the internal activity. The activity rate in the second half of the stimulus-response with the delayed gain control is uplifted compared to the absence of the gain control.}

\en{Figure \ref{fig:delayed_gain}\textbf{D} displays the dynamics of this stimulus-response in phase diagrams. The left panel shows a $U$-$\beta$ phase diagram. Since the internal activity involves neurons 1, 2, and 3, the initial stimulus-response of these neurons increases the internal activity $U$ that involves all neurons in the current definition. $U$ then decreases because the activity rates of neurons 1, 2, and 3 fall. Since $\beta$ is constant in the absence of the delayed gain control, this dynamics is represented by a line (Fig.~\ref{fig:delayed_gain}\textbf{D}, Left panel, dotted line). To the contrary, in the presence of the delayed gain control, the dynamics draw a cycle in the $U$-$\beta$ phase diagram (Fig.~\ref{fig:delayed_gain}\textbf{D}, Left panel, solid line). The reason is that the value of $\beta$ changes while $U$ varies in response to the stimulus. The panel on the right is a $V$-$\alpha$ phase diagram. Similarly, the dynamics are represented by a line in the absence of the delayed gain control (Fig.~\ref{fig:delayed_gain}\textbf{D}, Right panel, dotted line), and the dynamics draw a cycle in the presence of the delayed gain control (Fig.~\ref{fig:delayed_gain}\textbf{D} Right panel, solid line). By applying the first law of neural dynamics (Eq.~\ref{eq:first_law_dS_0}) to this stimulus-response cycle, we obtain the following equation:}
\begin{align}
0 = \oint \beta dU - \oint \alpha dV.
\label{first_law_cycle}
\end{align}
 \en{Note that entropy is a state variable. Since the states at the beginning and end of the cycle are the same, the left side of the above equation is $0$. On the other hand, the first and second terms on the right-hand side depend on the path of dynamics. The areas of the cycles in Fig.~\ref{fig:delayed_gain}\textbf{D} represent their magnitudes. Hence, Eq.~\ref{first_law_cycle} states that the two cycles in the $U$-$\beta$ phase diagram and $V$-$\alpha$ phase diagram have equal size for their areas.}

\en{One may consider that the size of the cycle quantifies, by the measure of entropy, the degree of modulation applied to the stimulus-response caused by the delayed gain control. In order to understand precisely what it means to have a large cycle as a neural dynamics, let us consider the following cycles. Figure \ref{fig:delayed_gain}\textbf{E} (Left panel) shows a phase diagram of $U$-$\beta$ when the minimum value of the gain control parameter $\beta$, which gives the maximum gain, is fixed at $0.9$. It displays several dynamics with different time constants $\tau_{\beta}$ of $\beta$. The larger the time constant, the larger the cycle. The reason is that $\beta$ takes a value close to 1 during the initial phase during which the internal activity increases, and $\beta$ becomes less than one during the phase in which the internal activity decreases, imposing the gain control. In this case, the initial response is not uplifted very much, whereas the response persists for a longer period during the descending phase (Fig.~\ref{fig:delayed_gain}\textbf{E} Right panel). Thus, the cycle magnitude indicates how much the changes in the internal activity contributed to sustaining the stimulus-response.}

\en{If the neural dynamics forms a cycle in the phase diagram of $U$-$\beta$ or $V$-$\alpha$, this neural dynamics works similarly to a heat engine in thermodynamics. In the case of a heat engine, a heat bath supplies heat to a gas, and then the gas emits the heat to a colder heat bath. The energy difference is taken out as the work. That is, the system produces the work by taking advantage of the temperature difference. Much in the same way, this neural dynamics acts as an information-theoretic engine that converts entropy changes due to the internal activity (the first term, $\Delta S^{\rm int}=\oint \beta dU $) into entropy changes of the stimulus-related activity (the second term, $\Delta S^{\rm ext}=\oint \alpha dV$). The difference in the parameter $\beta$ of the gain control causes the entropy change in the internal activity. We call the neural dynamics of stimulus-response subject to the time-delayed modulation \textbf{Neural Engine} \cite{shimazaki2015neurons,shimazaki2018neural}.}

\begin{figure}[!t]
\begin{center}
\includegraphics[width=.4\textwidth]{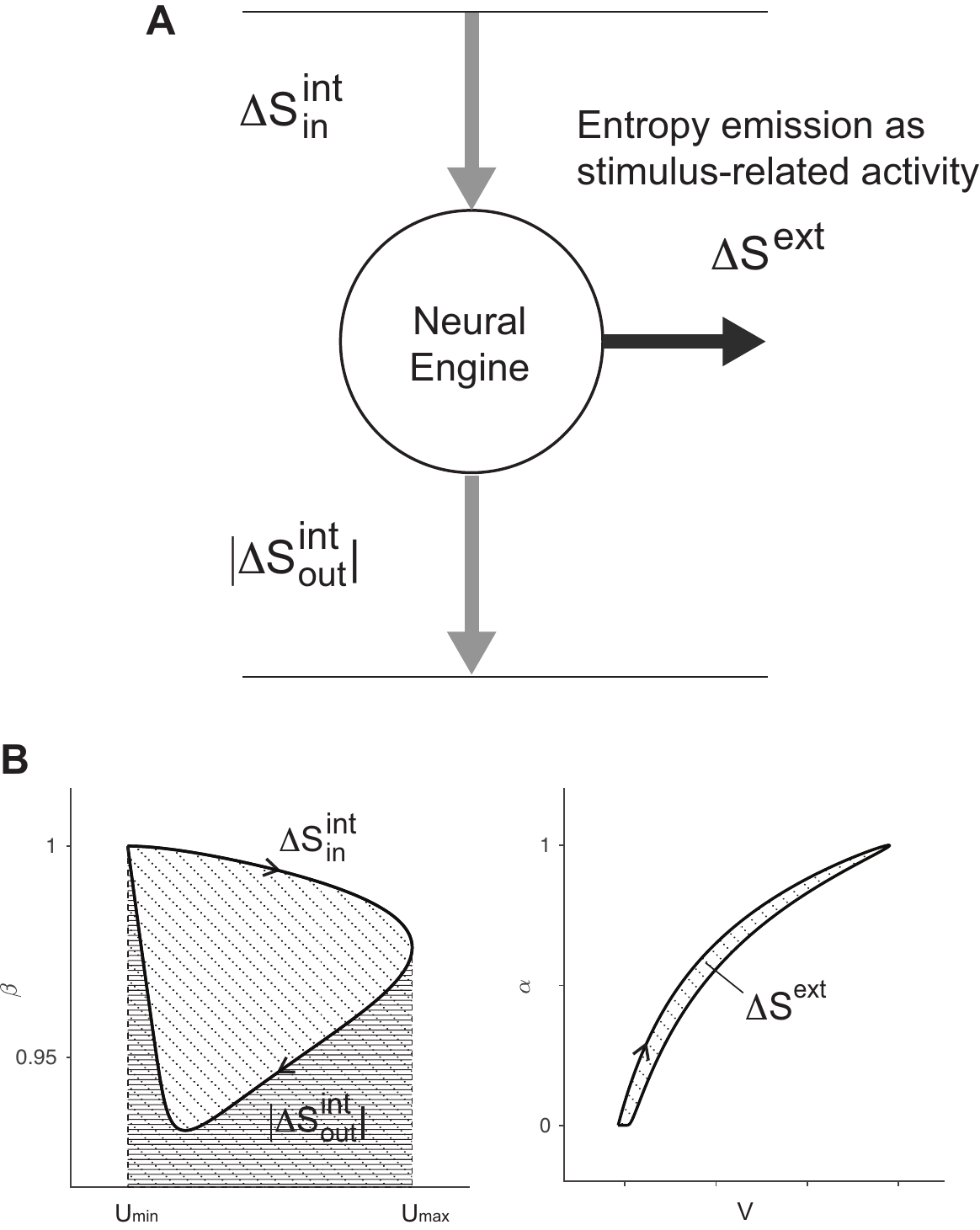}
\end{center}
\caption{ \en{Schematic diagrams of a neural engine. 
   \textbf{A} The difference between the increase in entropy of the neural activity (in-flow) caused by the internal activity during the upward phase of the response $\Delta S^{\rm int}_{\rm in}$ and the decrease in entropy (out-flow) during the downward phase $ |\Delta S^{\rm int}_{\rm out}| $ is discharged as an entropy change due to stimulus-related activity $\Delta S^{\rm ext}$.
   \textbf{B} Left panel: Entropy change by the internal activity. The increase in entropy in the ascending phase $\Delta S^{\rm int}_{\rm in}$ is represented by sloping lines, and the decrease in entropy in the descending phase $ |\Delta S^{\rm int}_{\rm out}| $ is marked by horizontal lines. Note that the sloping lines include the area of the horizontal lines. The shaded area extends to $\beta=0$ but is not shown. Right panel: Change in entropy due to the stimulus-related activity, $\Delta S^{\rm ext}$.
}
}
\label{fig:information_cycle_schematics}
\end{figure}

\en{Figure \ref{fig:information_cycle_schematics} shows schematic illustrations of the neural engine. Let the internal activity at the stimulus onset be ${U_{\rm min}}$. We define $\Delta S^{\rm int}_{\rm in}$ as the change in entropy of the neural activity (recognition model) caused by the internal activity until $U$ achieves the maximum $U_{\rm max}$. This entropy change is represented as the shaded area with sloping lines in the left panel of Figure \ref{fig:information_cycle_schematics}\textbf{B}, which is a positive value in this example. Next, let $\Delta S^{\rm int}_{\rm out}$ be the entropy change caused by the internal activity when $U$ returns to the level of spontaneous activity from its maximum value. The sign of $\Delta S^{\rm int}_{\rm out}$ is negative, and its magnitude is expressed as the area of the horizontal lines in the figure. The area of the cycle is computed as $\Delta S^{\rm int}=\Delta S^{\rm int}_{\rm in} - |\Delta S^{\rm int}_{\rm out}|$. This area is not zero if neurons experience the delayed gain control. In other words, changes in the internal activity increase the entropy of neural activity during the initial response to a stimulus (ascending phase) and decrease the entropy in the descending phase, but their magnitudes are different. As a result, a cycle of a single stimulus-response increases entropy by $\Delta S^{\rm int}$. The excess entropy produced by the internal activity is discharged as the entropy of the stimulus-related activity $\Delta S^{\rm int}_{\rm out}$ (Fig.~\ref{fig:information_cycle_schematics}, Right panel).} 

\en{It should be noted that, in order for the neural engine to emit the stimulus-related entropy, the feature for the internal activity $\mathcal{H}_0({\mathbf{x}})$ must have overlapping elements with the stimulus-related feature $\mathcal{H}_1({\mathbf{x}})$ (i.e., neurons 1, 2, and 3 in this example), otherwise $\beta$ and $\alpha$ independently control distinct populations. In such case, the neural dyamics is once again written by lines in the phase diagrams. It is this interaction between the features that makes it possible to realize the nonlinear gain control of the stimulus-realted activity by the internal activity \cite{shimazaki2018neural} (Fig.~\ref{fig:ising_example}), which results in exchange of the entropy when the modulation is delayed.}

\begin{figure*}[!t]
\begin{center}

\en{ \includegraphics[width=.9\textwidth]{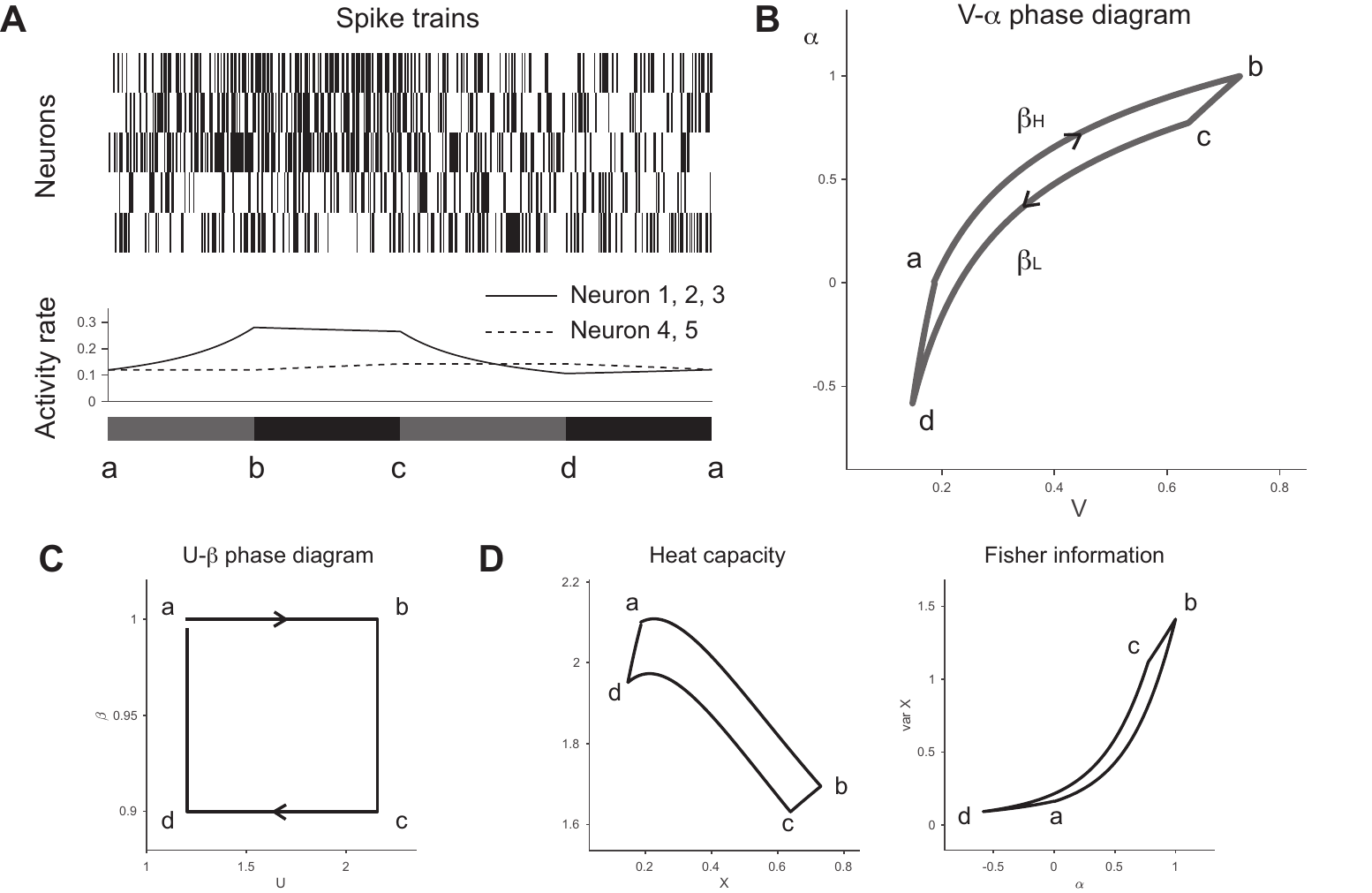} }
\end{center}
\caption{ \en{ The neural engine that achieves the highest entropic efficiency. The number of neurons and parameters for features are the same as in Fig.~\ref{fig:ising_example}.
   \textbf{A} Top panel: Exemplary spiking activity of 5 neurons. Lower panel: activity rates of neurons.
   \textbf{B} $V$-$\alpha$ phase diagram.
   \textbf{C} $U$-$\beta$ phase diagram.
   \textbf{D} Fluctuation of the neuronal population. Left panel: Fluctuation of the entire system (heat capacity). Right panel: Fluctuation of the stimulus-related activity (Fisher information about $\alpha$).
}
}
\label{fig:carnot_engine}
\end{figure*}

\en{The efficiency of retaining information can be quantified in the same way as the thermal efficiency of an engine. The following entropic efficiency was proposed to measure how much of the entropy increased by the internal activity during the initial response was used to change the stimulus-related entropy \cite{shimazaki2015neurons,shimazaki2018neural}:}
\begin{align}
\eta &\equiv \frac{ \Delta S^{\rm{ext}} } {\Delta S^{\rm{int}}_{\rm{in}} }  
	=   1 - \frac{| \Delta S^{\rm{int}}_{\rm{out}} |} {\Delta S^{\rm{int}}_{\rm{in}} }.  
	\label{efficiency} 
\end{align}

\en{We can consider the neural dynamics that yields the largest cycle, given minimum and maximum values of $\beta$ that provide maximum and minimum gain. Figure \ref{fig:carnot_engine} displays such a cycle. In this cycle, the stimulus-response is not modulated until it reaches its maximum value. Then the stimulus-response receives the maximum gain control after it starts to decrease. In the $U$-$\beta$ phase diagram, a rectangular cycle represents such a dynamics (Fig.~\ref{fig:carnot_engine}\textbf{C}). Since the gain control does not contribute to increase the initial response, rather, all of the gain modulation caused by the change in $\beta$ is used to prolong the stimulus-response, we consider it to be the most efficient dynamics retaining the stimulus-response. This ideal cycle is an analogue of the Carnot cycle that achieves the maximum thermal efficiency in thermodynamics.\footnote{The cycle that gives the highest entropic efficiency is not identical to the Carnot cycle that gives the highest thermal efficiency. The adiabatic processes in the Carnot cycle ($dS=0$) is replaced with constant internal activity ($dU= 0$) in the cycle that gives the maximum entropic efficiency. See \cite{shimazaki2015neurons,shimazaki2018neural} for the detailed description of this cycle.} The entropic efficiency of this ideal cycle is}
\begin{align}
	\eta_{e} = 1 - \frac{|\beta_L  (U_{\rm min}-U_{\rm max}) |} {\beta_H (U_{\rm max}-U_{\rm min}) } =  1 - \frac{\beta_L} {\beta_H }.
\end{align}	
 \en{This entropic efficiency is determined only by $\beta$. Here $\beta_H$ and $\beta_L$ are the $\beta$ during the upward and downward phase of the stimulus-response, respectively. We let $\beta_H=1$. The inset in the right panel of Fig.~\ref{fig:delayed_gain}\textbf{D} shows the entropic efficiency of the neural engine with different time constants, $\tau_\beta$. As the $\beta$'s time constant $\tau_\beta$ increases, the efficiency of retaining the stimulus information increases, and approaches the upper limit of $\eta_{e}=1-0.9/1=0.1$.}

\en{In summary, we showed that the neural dynamics that realizes the Bayesian inference constitutes an information-theoretic engine when the stimulus information is modulated by the prior with time-delay. Based on this paradigm, we introduced the method for quantifying the amount of internal modulation to keep the stimulus information, and calculating its efficiency.}

\en{ \subsection{The maximum entropy principle for recognition and learning}\label{sec:the_laws_of_neurodynamics} }

\en{Finally, we introduce an unified view on the recognition and learning by extending the Bayesian inference under the maximum entropy principle. Following this formulation, we introduce thermodynamic approaches to treat the neural dynamics of recognition and learning, using the laws of entropy for neural activity \cite{shimazaki2015neurons,shimazaki2018neural,shimazaki2019principles}. More specifically, we derive the first law of neural dynamics expressing the conservation law for the entropy of neural activity, and the second law for learning.}

\en{We construct a recognition model $q(\mathbf{x}|{\mathbf{Y}})$ as a probability mass/density function with no structure other than the ones made by the following constraints on the prior and observation model,}
\begin{eqnarray}
\langle - \log p({\mathbf{x}} |{\mathbf{w}}) \rangle &=& \mathcal{U}, \label{eq:constraint_prior} \\
\langle - \log p({\mathbf{Y}}|{\mathbf{x}}, {\mathbf{w}}) \rangle &=& {\mathcal{V}}.
\label{eq:constraint_observation}
\end{eqnarray}
 \en{Here, $\langle \cdot \rangle$ represents expectation by $q(\mathbf{x})$, and $\mathcal{U}$ and $\mathcal{V}$ are the expected values.\footnote{Note that definitions of $\mathcal{U}$ and $\mathcal{V}$ are different from $U$ and $V$ in the previous section. $\mathcal{U}$ and $\mathcal{V}$ include constant terms not related to $\mathbf{x}$. Also, note that the definitions of $V$ and $\mathcal{V}$ have opposite signs with respect to the stimulus-related feature.} This is an entropy maximization problem with constraints. We can solve this problem by the method of Lagrange multipliers to minimize the following functional $\mathcal{G}$:}
\begin{equation}
\mathcal{G}[q] = - S + \beta \langle - \log p({\mathbf{x}} |{\mathbf{w}}) \rangle + \alpha \langle - \log p({\mathbf{Y}}|{\mathbf{x}}, {\mathbf{w}}) \rangle,
\label{eq:free_energy}
\end{equation}
 \en{where $S$ is the entropy of the recognition model (Eq.~\ref{eq:entropy_recognition}). $\beta$ and $\alpha$ are constants called the Lagrange multipliers. They are chosen to satisfy the constraints given by Eqs.~\ref{eq:constraint_prior} and \ref{eq:constraint_observation}. In terms of $\mathcal{G}$, these constraints are expressed as the following formula:}
\begin{eqnarray}
\frac{\partial \mathcal{G}}{\partial \beta } &=& {\mathcal{U}} \label{eq:constraint_U_G}, \\
\frac{\partial \mathcal{G}}{\partial \alpha } &=& {\mathcal{V}} \label{eq:constraint_V_G}.
\end{eqnarray}
 \en{In the following, we call $\mathcal{G}$ \textbf{free energy}, and consider the problem of minimizing the free energy.}

\en{The probability mass/density function $q({\mathbf{x}}|{\mathbf{Y}})$ that minimizes the free energy $\mathcal{G}$ is given as the one that makes the following variation of $\mathcal{G}$ equals to zero,}
\begin{align}
\delta \mathcal{G}[q] &= \int \delta q({\mathbf{x}}|{\mathbf{Y}}) [1 + \log q(x) - \beta  \log p({\mathbf{x}} |{\mathbf{w}})  \nonumber\\
 &\phantom{=======} - \alpha  \log p({\mathbf{Y}}|{\mathbf{x}}, {\mathbf{w}}) ] \, d{\mathbf{x}}.
\label{eq:delta_free_energy}
\end{align}
 \en{Threfore, we obtain \cite{shimazaki2019principles}}
\begin{equation}
q({\mathbf{x}}|{\mathbf{Y}}) = \frac{1}{{{{\mathcal{Z}}_{\beta ,\alpha}}({\mathbf{Y}})}}{e^{ \beta  \log p({\mathbf{x}}|{\mathbf{w}} ) + \alpha \log p({\mathbf{Y}}|{\mathbf{x}},{\mathbf{w}} ) }}.
\label{eq:recognition_exp}
\end{equation}
 \en{Here we additionally constrained the function to be the probability density by introducing the normalization function, ${\mathcal{Z}}_{\beta, \alpha}$:}
\begin{equation}
{\mathcal{Z}}_{\beta, \alpha}({\mathbf{Y}}) = \sum_{\mathbf{x}} e^{ \beta  \log p({\mathbf{x}}|{\mathbf{w}} ) + \alpha \log p({\mathbf{Y}}|{\mathbf{x}},{\mathbf{w}} ) }.
\label{eq:normalization_max_ent}
\end{equation}
 \en{Note that this function is also called the partition function, and becomes the marginal likelihood function at $\beta=1$ and $\alpha=1$: ${\mathcal{Z}}_{\beta=1,\alpha=1} =  p({\mathbf{Y}}|\mathbf{w} )$. Since the above probability mass/density function was derived under the maxium entropy principle, it is called the maximum entropy model.}

\en{The recognition model obtained by the maximum entropy principle (Eq.~\ref{eq:recognition_exp}) extends the Bayesian formula, where $\beta$ and $\alpha$ represent weights of the prior and observation models in constructing the approximate posterior distribution. For example, we may employ a Gaussian distribution with the covariance matrix $\boldsymbol{\Sigma}$ as the observation model (Eq.~\ref{eq:observation_ising}), and the Ising model (Eq.~\ref{eq:prior_ising}) with the parameter $\boldsymbol{\Omega}$ as a prior distribution. Then the maximum entropy model (Eq.~\ref{eq:recognition_exp}) becomes the recognition model used to construct the neural engine in the previous section (Eq.~\ref{eq:approx_posterior_ising}) (but with a different normalization term).\footnote{In the recognition model used in the previous section, we used the Gaussian distribution with the covariance matrix $\alpha \boldsymbol{\Sigma}$ as the observation model (Eq.~\ref{eq:observation_ising}) and the Ising model with the parameter $\beta \boldsymbol{\Omega}$ as a prior distribution (Eq.~\ref{eq:prior_ising}). The scaling parameters $\beta$ and $\alpha$ in the observation model and prior distribution are the linear parameters of the features (functions of $\mathbf{x}$). We can take such parameters out of the observation and prior models, and treat them as linear weights of the standardized observation and prior models in the maximum entropy model. In general, if we use generalized linear models for the observation model and prior distribution, we may handle their linear parameters in the same way. However, the constant terms in the features that do not relate to $\mathbf{x}$ are different in these two approaches. For example, we have $\beta \psi(\boldsymbol{\Omega})$ as a constant term unrelated to $\mathbf{x}$ in the maximum entropy model (Eq.~\ref{eq:recognition_exp}), whereas we obtain $\psi(\beta\boldsymbol{\Omega})$ as the constant term in the recognition model used in the previous section (Eq.~\ref{eq:approx_posterior_ising}). The difference in these terms does not affect the distribution, but results in different normalization terms.} Thus, to repeat the previous section, if $\beta=1$ and $\alpha=1$, this recognition model is an exact posterior distribution. If $\beta=1$ and $\alpha=0$, it represents a spontaneous activity. Hence, the dynamics forming the optimal inference from the spontaneous activity during the stimulus presentation is realized by changing $\alpha$ from $0$ to $1$ under $\beta=1$.}

 \en{From Eq.~\ref{eq:entropy_recognition}, the entropy for the maximum entropy model (Eq.~\ref{eq:recognition_exp}) is given as}
\begin{align}
S &=  \beta \langle -\log p({\mathbf{x}} |{\mathbf{w}}) \rangle + \alpha \langle -\log p({\mathbf{Y}}|{\mathbf{x}}, {\mathbf{w}}) \rangle \nonumber\\
&\phantom{=====} + \log {\mathcal{Z}}_{\beta, \alpha}({\mathbf{Y}}).
\label{eq:entropy_legendre}
\end{align}
 \en{Thus, by inserting the above into the equation of the free energy (Eq.~\ref{eq:free_energy}), we obtain}
\begin{equation}
\mathcal{G}(\beta,\alpha) =  - \log {\mathcal{Z}}_{\beta,\alpha}({\mathbf{Y}}).
\end{equation}
 \en{That is, the free energy of the maximum entropy model obtained under the constraints (Eq.~\ref{eq:recognition_exp}) is given as the negative logarithm of its normalization (partition function) of the probability function. Note that the free energy $\mathcal{G}$ is no longer a functional because we specified the exponential form of the recognition model. Hence we represent it as a function with respect to the parameters of the recognition model, $\beta$ and $\alpha$. Here we call $\mathcal{G}(\beta,\alpha)$ as the \textbf{Gibbs free energy} Further, from Eqs.~\ref{eq:constraint_U_G} and \ref{eq:constraint_V_G}, the total derivative of $\mathcal{G}(\beta,\alpha)$ becomes}
\begin{eqnarray}
d{\mathcal{G}}(\beta,\alpha) &=& \left( \frac{{\partial {\mathcal{G}}}}{{\partial \beta}} \right)_{\alpha} d\beta +   \left( \frac{{\partial {\mathcal{G}}}}{{\partial \alpha}} \right)_{\beta} d\alpha \nonumber\\
&=& {\mathcal{U}} d\beta + {\mathcal{V}} d \alpha.
\label{eq:d_gibbs_free_energy}
\end{eqnarray}
 \en{}

\en{The learning of the external world by the generative model is achieved by changing the parameters ${\mathbf{w}}$ of the prior and observation model to maximize the marginal likelihood function. We note that the partition function ${\mathcal{Z}}_{\beta,\alpha}({\mathbf{Y}})$ becomes the marginal likelihood function at $\beta=1$ and $\alpha=1$. Therefore, decreasing the Gibbs free energy under the optimal inference is equivalent to learning according to maximization of the log marginal likelihood. We can express the dynamics of learning as follows.}
\begin{equation}
\frac{d\mathbf{w}}{dt} = - \epsilon \frac{\partial \mathcal{G}(\beta,\alpha) }{\partial \mathbf{w}},
\label{eq:learning_dynamics}
\end{equation}
 \en{where $\epsilon$ is a positive learning coefficient, and we impose that the learning carried out only if the parameters $\beta$ and $\alpha$ take in the vicinity of $\beta=1$ and $\alpha=1$. Here we note that Eq.~\ref{eq:d_gibbs_free_energy} dictates how the Gibbs free energy changes by changing the parameters, $\beta$ and $\alpha$, but does not include contributions by changing ${\mathbf{w}}$. Thus, if the learning following Eq.~\ref{eq:learning_dynamics} takes place, we expect}
\begin{equation}
d{\mathcal{G}}(\beta,\alpha) \leq {\mathcal{U}} d\beta + {\mathcal{V}} d \alpha.
\label{eq:dG}
\end{equation}
 \en{In particular, if the learning takes place while the parameters $\beta$ and $\alpha$ are fixed, we observe the contributions caused by the learning only: $d{\mathcal{G}} \leq 0$.}

 \en{Finally, we derive the conservation law for the entropy of neural activity, and show that decreasing the Gibbs free energy by learning is equivalent to increasing the entropy. From Eq.~\ref{eq:entropy_legendre}, the entropy of the recognition model is obtained as}
\begin{equation}
S( {\mathcal{U}}, {\mathcal{V}})  = \beta {\mathcal{U}} + \alpha {\mathcal{V}} - \mathcal{G}(\beta,\alpha).
\label{eq:entropy_legendre_G}
\end{equation}
 \en{The total derivative of the entropy is given as}
\begin{align}
dS({\mathcal{U}},{\mathcal{V}}) &= d(\beta {\mathcal{U}} + \alpha {\mathcal{V}}) - d\mathcal{G}(\beta,\alpha) \nonumber\\
&= (\beta d{\mathcal{U}} +  {\mathcal{U}} d\beta ) + (\alpha d{\mathcal{V}} + {\mathcal{V}} d\alpha) \nonumber\\
&\phantom{==========}- d\mathcal{G}(\beta,\alpha).
\label{eq:dS_expansion}
\end{align}
 \en{We obtain the following euqaiton by inserting the total derivative of the Gibbs free energy (Eq.~\ref{eq:d_gibbs_free_energy}) to this equation,}
\begin{align}
dS = \beta d{\mathcal{U}} + \alpha d{\mathcal{V}}. %
\end{align}
 \en{This is the first law of neural dynamics when we use the maximum entropy model as the neural activity. As mentioned in the previous section, it contains the same mathematical structure as the first law of thermodynamics that represents the conservation of heat and energy. For example, we can rewrite $\alpha$ as $\alpha=\beta f$ by introducing a parameter $f$ known as a force, and consider $\beta$ and $f$ (rather than $\beta$ and $\alpha$) to be independent parameters. In this case, $\beta$ is a parameter that controls both the prior and the observation model simultaneously. The parameter $\beta$ is then a natural parameter dually to the expected value, $\mathcal{E}=\mathcal{U}+f\mathcal{V}$, and becomes an analog of the inverse temperature that controls the energy of the system in thermodynamics. It is possible to consider the gain control by such $\beta$ that results in neurons' sensitivity change (see Appendix); however, in this article, we discussed the neural dynamics represented by $\beta$ and $\alpha$ as its independent variables.}

 \en{Next, if we apply the inequaility (Eq.~\ref{eq:dG}) obtained under the learning to Eq.~\ref{eq:dS_expansion}, we obtain}
\begin{align}
dS \geq \beta d{\mathcal{U}} + \alpha d{\mathcal{V}}. %
\end{align}
 \en{In particular, given that we fix $\beta$ and $\alpha$ ($U$ and $V$), this inequality represents the law of increasing entropy: $dS \geq 0$. That is, learning that follows the marginal likelihood maximization results in an increase in the entropy of neural activity. Following the second law of thermodynamics, we call this law \textbf{the second law of neurodynamics}.\footnote{Note that here we consider a quasistatic process. If we consider the irrevasible neural dyanmics expected for history-dependent models, we additionally obtain the entropy production caused by the causal effects \cite{crooks1999entropy,ito2013information}.} However, simply increasing the entropy of the recognition model or decreasing Gibbs free energy is not sufficient for learning. Rather, the organism needs to make optimal inference as much as possible by constructing a distribution close to the exact posterior distribution achieved at $\beta=1$ and $\alpha=1$.\footnote{The situation is the same as in the Expectation-Maximization alogrithm of machine learning. The parameter optimization (M-step) must come with the optimal inference (E-step). Otherwise, the difference between the marginal likelihood and its lower bound diverges.}}

\en{In this section, we introdced an extended framework of the Bayes' formula that integrates the prior distribution and observation model based on the maximum entropy principle, and derived laws for entropy on the dynamics of organism's learning and recognition.}

\en{ \section{Summary and prospects} }

\en{This article explained the dynamics of organisms' learning and recognition from the perspectives of the Bayesian inference and thermodynamics. Based on the experiments suggesting that the brain acquires a generative model of the world and the experiments revealing the temporal structure of recognition dynamics in living organisms, we modeled the stimulus-response activity of a spiking neural population that dynamically performs the Bayesian inference. We then explained the thermodynamic view of the Bayesian brain, introducing the laws for entropy of neural dynamics.  Notably, neural activity operates as an information-theoretic engine (neural engine) when the stimulus information is modulated by the delayed gain control. In the following, we discuss the mertis of treating the brain as the information-theoretic engine, and introduce the analysis framework to test the hypothesis from observed spiking data.}

\en{\textbf{Utilities of the delayed gain control} What are the benefits of the delayed gain control for organisms' learning and recognition? One feature of the delayed gain control is to keep the effect of the stimulus-response longer without excessively increasing the maximum values of the firing rates (See Fig.~\ref{fig:delayed_gain}\textbf{E}). It was the entropic efficiency that quantified this feature: The cycle with the highest entropic efficiency prolongs the stimulus-response without changing the maximum firing rates of neurons (Fig.~\ref{fig:carnot_engine}). Such neural activity has a clear biological benefit. By avoiding intensive firing happening in a short time, the network activity avoids undergoing an inactive state caused by synaptic depression (for example, due to depletion of synaptic vesicles or sensitization of receptors to ligands) \cite{curto2009simple}. It also has merits in inference and learning. With such dynamics, the network can construct a nearly optimal posterior distribution at an initial response, which allows the organisms to make an appropriate inference (Fig.~\ref{fig:information_cycle_schematics}\textbf{B}). The contextual information modulates the stimulus-response only after the networks made the optimal inference about the stimulus. As shown in Section \ref{sec:the_laws_of_neurodynamics}, learning needs to take place under the optimal posterior distribution. If biological learning based on neural spikes occurs when the neural system is highly active as expected for the long-term potentiation (LTP) or spike-timing-dependent plasticity (STDP), then the appropriate learning can naturally take place by the presence of this unbiased initial response.}

\en{\textbf{Hierarchy of temporal scales and its mechanism} The delayed gain control prolongs the stimulus-evoked activity in the early sensory areas. We may say that the early sensory cortex, which operates on a relatively fast time-scale, retains the sensory information for a more extended period with the help of the higher cortical areas that operate on a longer time-scale. Is such a temporal hierarchy observed in the brains? Ogawa et al. investigated the difference in time-scales between the area V4 and the frontal eye field (FEF), and found that the intrinsic time-scale of FEF is much longer than that of V4 \cite{ogawa2010differential}. They also pointed out that representations of V4 neurons dynamically change from visual features to behaviorally-relevant features, and that this may be caused by modulation by the feedback input from FEF \cite{ogawa2006neuronal}. Murray et al. reported that the intrinsic time-scale of cortical spiking activity slows down from the sensory cortex to the prefrontal cortex \cite{murray2014hierarchy}. In this study, they examined the intrinsic time-scales of six different cortical areas using the decay time of autocorrelation functions for spike counts during pre-trial periods. They found the intrinsic time-scale slows down from the sensory areas (MT, S1/S2) to the prefrontal area (LIP, LPFC, ACC) in the range of 50ms to 350 ms, where they observed the most extended time-scale at the anterior cingulate cortex (ACC). They also found a positive correlation between the time-scale of reward-induced firing-rate modulation and intrinsic time-scale in the prefrontal cortex, suggesting that the intrinsic time-scale is the basis of the dynamics of task-dependent neural activity. Although further experimental studies are nessesary to uncover the temporal hierachy in the brain, these studies support that the feedback inputs from higher cortical areas, in general, can exhibit longer time-scales.\footnote{From the perspective of adaptive systems, it is conceivable that the brain developed the intrinsic temporal hierarchy to capture a temporal hierarchy existing in natural stimuli. However, it will not be easy to test this idea empirically. As a constructive approach, researchers are studying whether recurrent neural networks with multiple intrinsic time-scales can acquire temporally hierarchical contexts \cite{yamashita2008emergence}.}}

\en{Each cortical area in the brain is characterized by distinct network architecture and cytoarchitecture defined by cell size, density, and layer structure. These architectural differences are thought to underlie the different time-scales and firing characteristics of cortical neurons \cite{murray2014hierarchy,mochizuki2016similarity}. For example, the series of histological studies by Elston and colleagues revealed that, in monkeys and humans, the dendrites of pyramidal cells extend wider spatial areas as we move from the visual to prefrontal cortices, and the spine density on the dendrites also increases along the pathway \cite{elston1998morphological,elston2002cortical,elston2003cortex}. Thus, they suggested that the neurons in the higher-order cortical areas integrate information more intensively than the neurons in early visual areas. Researchers argued that such differences in cellular architecture and morphology are behind the hierarchical time-scales in neural information processing, and that the fast time-scale of the visual cortex may be better suited to process rapidly changing visual stimuli \cite{elston2003cortex,maimon2009beyond,murray2014hierarchy}.}

\en{\textbf{Integrating the dynamics with different time-scales} The critical prerequisite that makes neural activity constitute the information-theoretic engine is the integration of dynamics operating with different time-scales. In this article, we provided significantly simplified neural dynamics governed by just two parameters with fast and slow time-scales for the observation and prior, to illustrate how neurons can dynamically realize the Bayesian inference. However, we cannot expect the two parameters to represent the actual activity of a neural population. We need advanced statistical modeling methods to analyze empirically observed stimulus-evoked activity (see below for data analysis methods).}

\en{Moreover, it is not necessary to integrate observations and prior distributions with time-delay to constitute a narrowly-defined information-theoretic engine. For example, in Section \ref{sec:neuralengine}, we assumed the same time-scale for the activity rates of the three neurons encoding the stimulus. However, these neurons may respond to a stimulus with unique time-scales. In this case, the dynamics forms a cycle in the phase space without modulation caused by a change in the prior distribution. Nevertheless, the total entropy emission from the internal activity is zero. When quantifying the delayed modulation caused by awareness or attention, it would be important to assess if the cycle of stimulus-response increases by such a factor, resulting in larger entropy emission of the stimulus-related activity.}

\begin{figure*}[t]
\begin{center}
\includegraphics[width=.9\textwidth]{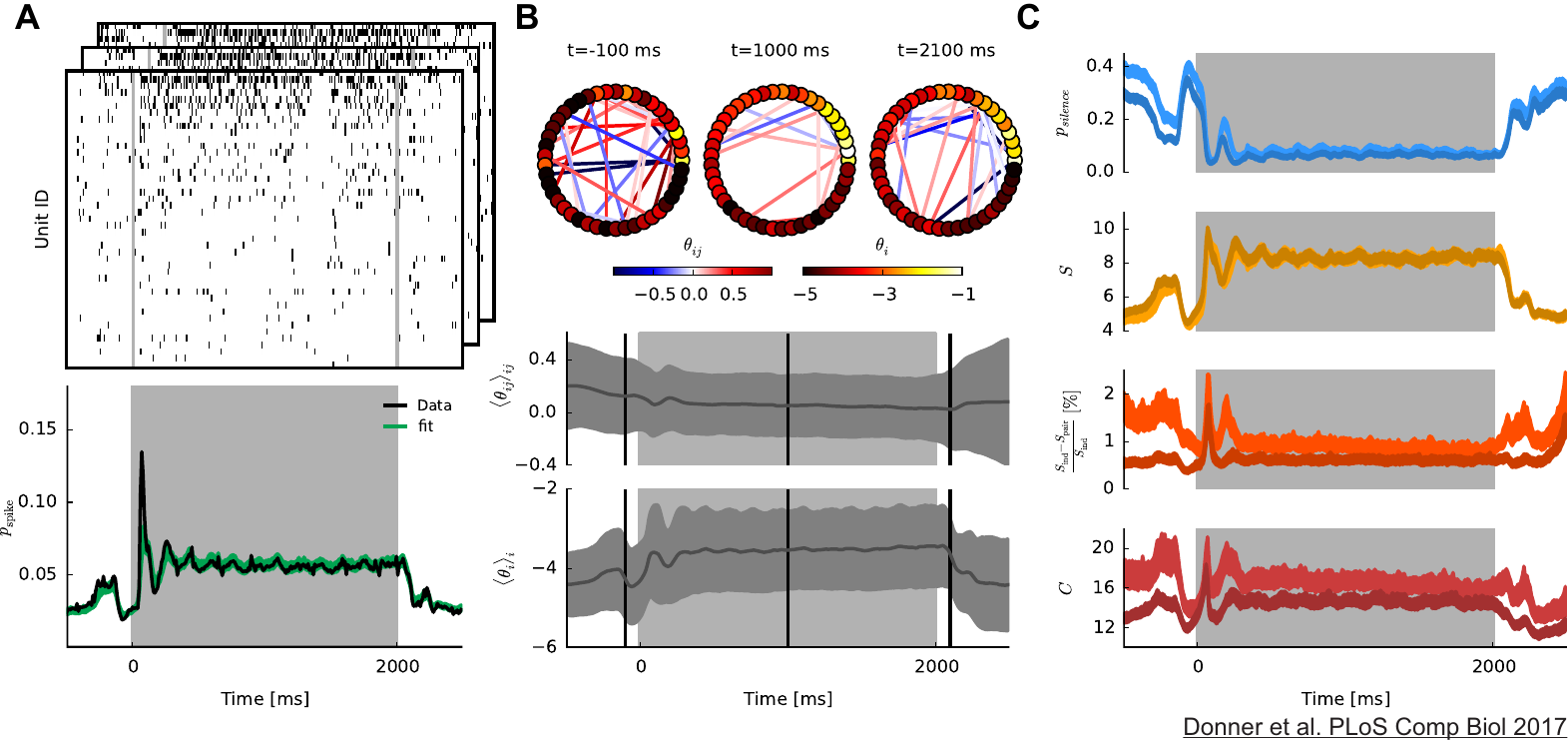}
\end{center}
\caption{
    \en{Analysis of the stimulus-evoked activity of monkey V1 neurons using a state-space Ising model. 
   \textbf{A} Top: Simultaneously recorded spike sequences of 45 neurons when a grating stimulus is presented. Bottom: Average firing probability of all neurons (data in black, model fit in green). The gray area indicates the period of stimulus presentation.
   \textbf{B} Top: A snapshot of the parameters of the estimated Ising model. Bottom: mean and standard deviation of the estimated parameters.
   \textbf{C} Dynamics of estimated thermodynamic quantities.  From top to bottom:  The credible intervals of probability of simultaneous silence, entropy, entropy ratio for pairwise correlations, and heat capacity. The darker credible intervals are the results of uncorrelated surrogate data obtained by trial shuffling.} 
   CC-BY Christian Donner, Klaus Obermayer, and Hideaki Shimazaki. Approximate inference for time-varying interactions and macroscopic dynamics of neural populations. PLoS computational biology, 13(1):e1005309, 2017. 
}
\label{fig:donner_ploscb_fig5}
\end{figure*}

\en{\textbf{Analysis of emprical data} The neural engine offers a way to assess a magnitude of the delayed modulation on neural activity thought to be involved in perceptual experience. Although modulation effects may be seen in the firing rates of individual neurons as reported in the literature, the view as a neural engine allows us to quantify its magnitude by entropy, a scalar macroscopic value of the system, while taking neurons' correlations into account. The question is then whether we can fit the Ising model to the observed time-series of spike data during the stimulus-response activity. More specifically, can we estimate dynamics of the time-varying distribution of the neural spiking activity expected during stimulus presentation, and quantify the system using the change of entropy? For this goal, the author and colleagues have been developing the state-space Ising models and its estimation method over the past decade \cite{shimazaki2009state,shimazaki2012state,shimazaki2013single,donner2017approximate,gaudreault2018state,gaudreault2019online} (Fig.~\ref{fig:donner_ploscb_fig5}).}

\en{In the state-space Ising model, we assume that the parameters of the Ising model (see Eq.~\ref{eq:prior_ising}) varies in time. The framework of data anlysis with this model is based on the sequential Bayesian algorithm for estimating the time-dependent Ising parameters and optimization of hyperparameters that dictate dynamics of the parameters. The method makes it possible to trace the dynamics of interactions between neurons, taking into account the time-dependent firing rates of individual neurons (Fig.~\ref{fig:donner_ploscb_fig5}\textbf{A, B}).  In addition, the method allows us to visualize the dynamics of macroscopic quantities of the system such as free energy, entropy, entropy related to correlated activities, heat capacity expressing system's fluctuation (Fig.~\ref{fig:donner_ploscb_fig5}\textbf{C}).}

\en{In general, the computational complexity in analyzing the patterns of population activity increases exponentially as the number of neurons increases. Analysis using the Ising model is no exception. However, many approximation methods have been developed for this classical model, including the pseudo-likelihood, mean-field, and Bethe approximations. By incorporating these methods into the state-space model, we estimated the parameters of a relatively large number of neurons ($\approx$50) and estimate the macroscopic quantities that characterize the system \cite{donner2017approximate}. Further, using the model of neural activity shown in Eq.~\ref{eq:recognition_ising}, Gaudreault et al.  proposed a method to estimate the time-dependent natural parameters while simultaneously learning the parameters of features \cite{gaudreault2019online}. Based on these recent adavances of the state-space Ising models, it is, in principle, possible to statistically fit the neural activity to the model of a neural engine to quantify the effect of feedback modulation, and relate it to perceptual experiences of animals.}

\en{\textbf{Quantifying consciousness} Is it possible to quantify consciousness by treating the brain as an information-theoretic engine? The entropy measure of the neural engine quantifies the retention of the stimulus information by the delayed gain control. The key factors underpinning this measure are a neuron's nonlinear response function (the basis of neural computation), the gain control using the nonlinearity (information integration), and recurrent inputs with a time-delay. Can we quantify consciousness with such a simple metric? Whether this is a reasonable measure depends on how much these underlying neuronal mechanisms are involved in conscious experiences, and, as a result, how closely the late component of stimulus-response activity is related to the conscious experiences or the level of consciousness of human subjects \cite{dehaene2003neuronal,mashour2014top}.}

\en{There are a handful of human studies on this issue with invasive electrophysiological recordings, but these supported the close relationship. Libet et al. examined subdural electrode recordings in the human somatosensory cortex and found that the initial response component was insufficient for a conscious experience of a tactile stimulus. They reported that the presence or absence of the subsequent late component correlated with the conscious experience \cite{libet1967responses}. The temporal characteristics of the disappearance of conscious experience by backward masking (in particular by common-onset masking) \cite{enns1997object,enns2000s} or transcranial magnetic stimulation (TMS) \cite{luck2000event} for humans also suggest that the late component is involved in the conscious experience.}

\en{Neural signals measured from humans often take the form of continuous signals such as EEG recordings or fMRI data with a few exceptions of electrophysiological recordings of spikes from patients with epilepsy. However, assessment of the neural signals as an information-theoretic engine is not limited to the model for time-series of spike trains discussed in this article. Instead, it can be constructed using models following an exponential family distribution. Therefore, it would be possible to assess if the EEG or fMRI signals recorded from humans works as a neural engine, and test if the late component of the stimulus-response is closely related to conscious experiences of humans, using the suggested entropy measure. More precisely, the locus ceruleus-norepinephrine (LC-NE) system is thought to regulate the level of animal's arousal, and it was suggested that activation of the LC-NE system acts as a brain-wide gain modulation \cite{aston2005integrative,eldar2013effects,donner2013brain}. Since the LC activity is tightly correlated with pupil diameters \cite{aston2005integrative,costa2016more}, it will be an important topic to study the relation between changes in pupil diameters and the suggested entropy measure for the delayed modulation, togather with subjective reports on the conscious experiences.}

\en{This article explained the neural dynamics of learning and recognition from the perspectives of the Bayesian inference and thermodynamics. We demonstrated that the stimulus-response subject to the delayed gain control is described as the dynamics of the Bayesian inference. We further introduced the thermodynamic view on this process. When the gain control takes place to integrate neural dynamics with various time-scales, the brain works as an information-theoretic engine. This new paradigm offers a unified computational and statistical view on the neural dynamics, providing quantative approaches to assess the organism's perceptual capacity from neural signals.}

\subsection*{acknowledgement}

\en{The author thanks Yusuke Hayashi, Fernando Rosas, Seiji Hirai for valuable comments on the manuscript. The author also thanks Tadashi Ogawa, Takatsune Kumada, and Masanori Murayama for sharing neurophysiological literature on the delayed modulation.}

\bibliographystyle{ieeetr}
\bibliography{references,statespaceising,neuralengine}

\newpage

\appendix

\en{ \subsection*{Appendix: The nonlinear response function, gain control, and generative model}\label{sec:learning_single_neuron} }

 \en{In this Appendix, we describe stimulus-evoked activity and the gain control using a generative model that is made of a population of independent neurons. Let $N$ be the number of neurons, and $x_i$ $(i=1,\ldots,N)$ be a binary variable taking $1$ and $0$, where $1$ denotes spiking and $0$ denotes no spikes. The probability that $x_i$ becomes $1$ is given by $\eta_i$. Hence the probability that $x_i$ becomes $0$ is given by $1-\eta_i$. We model how $\eta_i$, the expectation of $x_i$, changes according to an external stimulus. }

 \en{Let the stimulus presented to the animal be $\mathbf{Y}$. In the case of visual stimuli, $\mathbf{Y}$ is a column vector ($d\times 1$) of the pixel intensities of a grayscale image. We represent the expected value $\eta_i$ of the $i$th neuronal activity by the nonlinear response function to the input stimulus, $\eta_i=f_i(\mathbf{Y})$. For a binary random variable, the range must be restricted within 0 and 1. The function $f_i(\mathbf{Y})$ may be acquired through learning, but here we consider a logistic function as a nonlinear response function:}
\begin{equation}
f_i(\mathbf{Y})  = \frac{1}{{1 + {e^{ - (\varphi_{i0} + {\mathbf{Y}}' \boldsymbol{\varphi}_i)}}}},
\label{eq:response_fucntion_logistic}
\end{equation}
 \en{where $\boldsymbol{\varphi}_i$ is a $d$-dimensional column vector that represents a receptive field of the $i$th neuron. $\varphi_{i0} $ is a bias parameter, and adjusting this parameter results in the gain control. $\mu_i=\varphi_{i0} + {\mathbf{Y}}' \boldsymbol{\varphi}_i$ is called a linear predictor. }

 \en{The motivation to choose the logistic function as a response function does not come from biological plausibility but rather comes from its simplicity in statitical modeling, which originates from our choice of the Kullback-Leibler divergence as a metric for the probability space. The probability distribution of independent neurons is represented by the product of Bernoulli distributions:}
\begin{align}
p({\mathbf{x}}|{\mathbf{Y}}) &= \prod_{i=1}^{N} {\eta_i^{x_i}}{(1 - \eta_i )^{1 - x_i}} \nonumber\\ 
   &= \prod_{i=1}^{N} \exp [\log \frac{\eta_i }{{1 - \eta_i }} \cdot x_i + \log (1 - \eta_i )]. 
\end{align}
 \en{Based on the second equality, the Bernoulli distribution representing the $i$th neuronal activity belongs to an exponential familiy distribution: $f(x_i \mid \eta_i )=h(x_i)\exp \left[ \theta_i (\eta_i )\cdot T(x_i)-A(\eta_i )\right]$. Here $\theta_i$ is a linear parameter with respect to the feature $T(x_i)$, and we call them as natural/canonical parameters. In the case of a Bernoulli distribution, we obtain $\theta_i(\eta_i)=\log{\eta_i}/{(1-\eta_i)}$, which is called the logit function. The logit function is a natural link function of the Bernoulli distribution. Since the inverse of the logit function is a logistic function, $\theta_i$ is the linear predictor $\mu_i$. That is, if we represent the activity of independent neuronal populations as an exponential family distribution, we obtain}
\begin{align}
p({\mathbf{x}}| {\mathbf{Y}}) &= \prod_{i=1}^{N} \exp [ (\varphi_{i0} + {\mathbf{Y}}' \boldsymbol{\varphi}_i) x_i  \nonumber\\
&\phantom{====} - \log (1 + e^{\beta(\varphi_{i0} + {\mathbf{Y}}' \boldsymbol{\varphi}_i) })]. 
\label{eq:posterior_independent}
\end{align}
 \en{This equaition is seen as a regression of data on neuronal activity using a generalized linear model with a logit link function.}

\en{Next, we use the generative model to construct a posterior distribution of neural activity to represent the responses of the independent neurons described above. By doing so, we elucidate the relationship of the response function and its gain control with the generative model. First, we consider an independent Bernoulli distribution as a prior of the neural activity.}
\begin{equation}
p_{\beta}({\mathbf{x}}|{\boldsymbol{\omega}}) = \prod_{i=1}^{N} {\operatorname{e} ^{\beta \omega_i x - \log (1+ \operatorname{e} ^{\beta \omega_i} )}}.
\end{equation}
 \en{Namely, the activity rate of the $i$th neuron is given as $1/(1+e^{-\beta \omega_i})$ under this prior distribution. $\beta$ is a parameter that is common to all neurons. Next, we consider the observation model using a normal distribution.}
\begin{equation}
p_{\alpha}({\mathbf{Y}}|{\mathbf{x}},{\boldsymbol{\Phi}}) = \left(\frac{\alpha}{2\pi}\right)^{1/2} {\operatorname{e} ^{ - \frac{\alpha}{2}({\mathbf{Y}} - {\boldsymbol{\Phi}} {\mathbf{x}})'({\mathbf{Y}} -  {\boldsymbol{\Phi}}  {\mathbf{x}})}}.
\end{equation}
 \en{Here, for simplicity, we used a diagonal matrix as a covariance matrix whose variance is given by $\alpha^{-1}$. ${\boldsymbol{\Phi}}=[{\boldsymbol{\phi}_1},{\boldsymbol{\phi}_2},\ldots,{\boldsymbol{\phi}_N}]$ is a $d \times N$ dimensional matrix. The $i$th colmun ${\boldsymbol{\phi}_i}$ is a $d$-dimensional column vector, representing a stimulus by the $i$th neuron. It is called a basis vector or projection field. Here we assume that the basis vectors are orthogonal to each other, and their norm is 1: ${\boldsymbol{\Phi}}'{\boldsymbol{\Phi}}=\mathbf{I}$.}

 \en{The posterior distribution based on the generative model consisting of the prior distribution and observation model is expressed as follows.}
\begin{align*}
   &p_{\beta,\alpha}({\mathbf{x}}|{\mathbf{Y}}, \boldsymbol{\omega}, \boldsymbol{\Phi})= \frac{{p_{\beta}({\mathbf{x}}|\boldsymbol{\omega}) p_{\alpha}({\mathbf{Y}}|{\mathbf{x}},{\boldsymbol{\Phi}} )}} {{p_{\beta,\alpha}({\mathbf{Y}}|\boldsymbol{\omega}, \boldsymbol{\Phi})}} \nonumber\\ 
   &\propto \exp \left[ { \beta \sum_{i=1}^{N} \omega_i x_i  - \frac{\alpha}{2}({\mathbf{Y}} - {\boldsymbol{\Phi}} {\mathbf{x}})'({\mathbf{Y}} - {\boldsymbol{\Phi}} {\mathbf{x}}) } \right] 
\end{align*}
\begin{align}
   &\propto \exp \left[ {\beta \sum_{i=1}^{N} \omega_i x_i  + \alpha {\mathbf{Y}}' {\boldsymbol{\Phi}} {\mathbf{x}}  - \frac{\alpha}{2}{\mathbf{x}}' {\boldsymbol{\Phi}}' {\boldsymbol{\Phi}}{\mathbf{x}}} \right] \nonumber\\ 
   &\propto \prod_{i=1}^{N} \exp \left[ {(\beta \omega_i   - \frac{\alpha}{2} + \alpha {\mathbf{Y}}' {\boldsymbol{\phi}}_i) x_i} \right]. 
\label{eq:posterior_single}
\end{align}
 \en{Here we used ${\mathbf{x}}' {\boldsymbol{\Phi}}' {\boldsymbol{\Phi}}{\mathbf{x}}={\mathbf{x}}' {\mathbf{x}}=\sum_{i=1}^{N} x_i^2 =\sum_{i=1}^{N} x_i $. If we compare the neural activity derived as a posterior distribution (Eq.\ref{eq:posterior_single}) with the neural activity derived from the response function (Eq.~\ref{eq:posterior_independent}), we find that the posterior distribution follows the Bernoulli distribution whose expectation is given by the logistic response function, where the receptive field is ${\boldsymbol{\varphi}_i}=\alpha {\boldsymbol{\phi}}_i$, and a bias term is $\varphi_{i0}= \beta \omega_i-\alpha/2$ (Eq.\ref{eq:response_fucntion_logistic}).  Thus, changes in the parameters of the prior distribution, $\beta \omega_i$, modifies the bias term $\varphi_{i0}$, which results in the gain control of the stimulus-reponse function. In sum, using a generative model, we can model the stochastic neural activity that follows the logistic response function, and changes in the prior distribution can lead to the adaptation by the gain control.}

 \en{Further, if we define $\alpha=\beta f$ by introducing a new independent parameter $f$, then $\beta$ becomes a parameter that controls the prior and observation model simultaneously. In this case, we have}
\begin{align}
   &p_{\beta,\alpha}({\mathbf{x}}|{\mathbf{Y}}, \boldsymbol{\omega}, \boldsymbol{\Phi}) \nonumber \\
   &\propto \prod_{i=1}^{N} \exp \left[ {\beta (\omega_i   - \frac{f}{2} + f {\mathbf{Y}}' {\boldsymbol{\phi}}_i) x_i} \right].
\label{eq:posterior_single2}
\end{align}
 \en{Hence, by letting the bias term be $\varphi_{i0}=  \omega_i-f/2$ and the receptive field be ${\boldsymbol{\varphi}_i}=f{\boldsymbol{\phi}_i}$, the expected activity rate of the $i$th neuron is given as}
\begin{equation}
f_i(\mathbf{Y})  = \frac{1}{{1 + {e^{ - \beta (\varphi_{i0} + {\mathbf{Y}}' \boldsymbol{\varphi}_i)}}}}.
\label{eq:response_fucntion_logistic_beta}
\end{equation}
 \en{In this case, $\beta$ becomes the parameter that determines the slope of the curve of the logistic function. The higher the absolute value of $\beta$, the steeper the function becomes around $\mu_i=0$. Thus, by changing $\beta$ we can achieve the gain control that changes only the sensitivity of the response function to the stimulus without shifting the response function horizontally. This is a different type of gain control than that realized by changing the bias term, and can control the stochasticity of the response (For example, the response is deterministic if $\beta=\infty$, and random if $\beta=0$). Therefore, $\beta$ works analogouly to an inverse temperature of statistical physics.}

\end{document}